\documentclass[12pt, draftclsnofoot, onecolumn]{IEEEtran}
\usepackage{algpseudocode}
\usepackage{algorithm}
\usepackage{bm}
\usepackage{amsfonts}
\usepackage{amssymb}
\usepackage{amscd}
\usepackage{graphics}
\usepackage{xcolor}
\usepackage[dvips]{graphicx}
\usepackage{epsfig}
\usepackage{epstopdf}
\usepackage{subfigure}
\usepackage[font=scriptsize]{caption}
\usepackage[cmex10]{amsmath}
\usepackage{array}
\usepackage{eqparbox}
\usepackage{fancyhdr}
\usepackage{enumerate}
\usepackage{stackrel}
\usepackage{pdflscape}
\usepackage{longtable}
\usepackage{lineno}
\usepackage{mathtools}
\usepackage[noadjust]{cite}
\usepackage[export]{adjustbox}
\usepackage{amsmath}

\usepackage{latexsym}
\usepackage{cite,caption}
\usepackage{amsmath}
\usepackage{amssymb}
\usepackage{float}
\usepackage{epsfig}
\usepackage{subfigure,epstopdf,graphicx,array,algorithm}
\usepackage[]{graphicx}
\usepackage{algpseudocode}
\usepackage{textcomp}
\usepackage{cuted}

\newtheorem{defn}{\textbf{Definition}}
\newtheorem{theorem}{Theorem}[section]

\newcommand{\norm}[1]{\Vert#1\Vert}
\newcommand{\abs}[1]{\left\vert#1\right\vert}

\begin{document}
	
	\title{Online Learning to Cache and Recommend in the Next Generation Cellular Networks}
	
	\author{S.~Krishnendu, B. N. ~Bharath,
		and~Vimal~Bhatia 
		\IEEEcompsocitemizethanks{\IEEEcompsocthanksitem S. Krishnendu is with the Department of Electrical Engineering, Indian Institute of Technology Indore, India. e-mail: phd1701102001@iiti.ac.in. B. N. Bharath is with Department of Electrical Engineering, Indian Institute of Technology Dharwad, India. e-mail: bharathbn@iitdh.ac.in. Vimal Bhatia is with Department of Electrical Engineering, Center for Advanced Electronics, Indian Institute of Technology Indore, India and also with Faculty of Informatics and Management, University of Hradec Kralove. e-mail: vbhatia@iiti.ac.in \protect
		}
		\thanks{A part of this work has been accepted at VTC-2021 Spring conference \cite{krishnendu_bayes}.}
	}
	
	\maketitle
	\vspace{-3cm}
	\begin{abstract}
		An efficient caching can be achieved by predicting the popularity of the files accurately. It is well known that the popularity of a file can be nudged by using recommendation, and hence it can be estimated accurately leading to an efficient caching strategy. Motivated by this, in this paper, we consider the problem of joint caching and recommendation in a $5$G and beyond heterogeneous network. We model the influence of recommendation on demands by a Probability Transition Matrix (PTM). The proposed framework consists of estimating the PTM and use them to jointly recommend and cache the files. In particular, this paper considers two estimation methods namely a) \texttt{Bayesian estimation} and b) a genie aided \texttt{Point estimation}. An approximate high probability bound on the regret of both the estimation methods are provided. Using this result, we show that the approximate regret achieved by the genie aided \texttt{Point estimation} approach is $\mathcal{O}(T^{2/3} \sqrt{\log T})$ while the \texttt{Bayesian estimation} method achieves a much better scaling of $\mathcal{O}(\sqrt{T})$. These results are extended to a heterogeneous network consisting of $M$ small base stations (sBSs) with a central macro base station. The estimates are available at multiple sBSs, and are combined using appropriate weights. Insights on the choice of these weights are provided by using the derived approximate regret bound in the multiple sBS case. Finally, simulation results confirm the superiority of the proposed algorithms in terms of average cache hit rate, delay and throughput.
		
	\end{abstract}
	

	\IEEEdisplaynontitleabstractindextext

	%
	\IEEEpeerreviewmaketitle
	
	\vspace{-1cm}
	
	\section{Introduction}\label{sec:introduction}
	\IEEEPARstart{D}{ue} to the fast development of communication based applications, it is expected that there will be $5.3$ billion total Internet users ($66$ percent of global population) by $2023$, up from $3.9$ billion ($51$ percent of global population) in $2018$ \cite{Cisco18}. To enhance the users' Quality of Experience (QoE), several architecture have been proposed such as FoG networks, mobile edge computing (MEC) etc. These solutions enable the network to proactively predict the future content requests and store popular files closer to the edge devices. This reduces the delay and alleviates the backhaul congestion \cite{ywang23,zfeng,krishnendu2021fl, Krishnendu2019, araf23}. On the other hand rapidly growing file sizes, reduced cache sizes (compared to the traditional content delivery networks), and unpredictable user demands make the task of caching algorithms even more difficult. For example, the total data generated by Google per day is in the order of PBs, while installing 1TB memory in every small cell in the heterogeneous network will only shift less than 1 \% of the data for even one content provider. To overcome these issues, it has been observed that the user demands are increasingly driven by recommendation based systems. Recommendation based on an individual's preference have become an integral part of e-commerce, entertainment and other applications. The success of recommender systems in Netflix and Youtube shows that 80\% of hours streamed at Netflix, and 30\% of the overall videos viewed owes to recommender systems \cite{GomezUribe, Zhou}. With recommendation, the user's request can be nudged towards locally cached contents, and hence resulting in lower access cost and latency \cite{chatzieleftheriou2017caching}. This core concept was further expanded to encompass cache-assisted small cell networks in \cite{chatzieleftheriou2019joint}. Subsequently, numerous studies were carried out using diverse performance metrics to examine the combined recommendation and cache optimization \cite{yang2021,Sermpezis2018,wang2017,liu2018}.
	
	
	The recent success of integration of artificial intelligence in the wireless communications has further led to better understanding of user's behavior and the characteristics of the network \cite{hzhu}. Especially the edge networks can now predict the content popularity profile hence increasing the average cache hit. The high accuracy in prediction by the neural networks has resulted in many of the content popularity prediction models, such as, collaborative filtering with recurrent neural networks \cite{robin}, the stack auto-encoder \cite{liu}, deep neural networks \cite{chang}, and others. However, the local content popularity profile need not match the global prediction by the central server. Many of the recent works have proposed the edge caching strategies by learning the user preferences and content popularity \cite{muller,saputra19}. Context awareness helps in classifying the environment, hence enabling the intelligent decisions at the edge to select the appropriate content, for instance, Chen \textit{et al.} \cite{ychen} presented the edge cooperative strategy based on neural collaborative filtering. In \cite{araf23}, a context-aware caching policy through a cooperative Deep Reinforcement Learning-based algorithm is proposed. In \cite{qiao20}, the authors jointly optimize the content placement and content delivery in the vehicular edge computing and networks. The authors in \cite{heying18}, devised a novel integrated framework enabling the dynamic orchestration of networking, caching, and computing resources, thereby enhancing the performance of future vehicular networks. Jiang \textit{et al.} \cite{yjiang} used the offline user preferences data and statistical traffic patterns and proposed an online content popularity tracking algorithm. Nevertheless, the availability of offline data cannot always be guaranteed. Therefore, these studies make the assumption that users possess identical preferences and that there is no correlation among the data, even though this may not adhere to real-world scenarios.
	
	Based on the observation that the users demands (and hence better prediction of demands) can be nudged based on recommendations, in this work, we consider the problem of jointly optimizing recommendation and caching in a $5$G and beyond heterogeneous network consisting of Macro Base Stations (MBS), small base stations (sBSs) and users. The influence of recommendation on caching is modelled using a Probability Transition Matrix (PTM). Thus, one can optimize the recommendation to steer the requests in a way that results in a good cache hit performance. This can be done if each sBS has access to the PTM. Unfortunately, the PTMs are unknown and hence needs to be estimated. Towards this, we propose two estimation methods namely \texttt{Point estimation} and \texttt{Bayesian estimation}. The \texttt{Point} estimation method assumes that a random set of files will be recommended in the first $t$ time slots and the estimation is done using the frequency of occurrence of requests for each file conditioned on the recommended files. Being a naive method, the \texttt{Point} estimation method will be used to benchmark the \texttt{Bayesian} estimation method. In the \texttt{Bayesian} estimation scheme, the probability is estimated using a Bayesian approach, i.e., each row of the PTM is sampled from a Dirichlet distribution whose parameters are the naive estimates of the conditional probabilities (similar to the \texttt{point} estimation method). This method enables a nice balance between exploration and exploitation of caching and recommendation. For the above two methods, we provide the following results
	\begin{itemize}
		\item  For both the methods, a high probability guarantee on the estimated caching and recommendation strategies is provided. Irrespective of the estimation method, it is shown that with a probability of at least $1-\delta$, the performance of the proposed caching and recommendation strategy is $\epsilon$ close to the optimal solution. 
		\item An approximate high probability bound on the regret for \texttt{Bayesian estimation} method is provided. To compare and contrast the obtained regret bound, we also derive an approximate regret bound on a more powerful genie aided scenario using the \texttt{Point estimation} method.\footnote{For the explanation of the genie aided \texttt{point estimation} method, please refer to \ref{sec:theoretical_guarant}.} These regret bounds are shown to be data dependent. Hence, in order to get better insights, we carry out experiments to determine the scaling of the data dependent term of the regret. Using this result, we show that the approximate regret achieved by the genie aided \texttt{Point estimation} approach is $\mathcal{O}(T^{2/3} \sqrt{\log T})$ while the \texttt{Bayesian estimation} method achieves a much better scaling of $\mathcal{O}(\sqrt{T})$. 
		\item The above results are extended to a heterogeneous network consisting of $M$ sBSs with a central MBS. Since the estimates are available at multiple sBSs, it is possible to combine them at each sBS separately to obtain a better estimate. However, it is important to figure out the right weights to be used for different sBS. Towards this, we prove a regret bound for both the estimation methods. As time $T$ increases, each sBS collects more samples. Intuitively, as $T$ increases weights allocated to different sBSs' estimates should go down to zero. We confirm this intuition by using the regret bound that we derive. Further, we also show a scaling similar to the single sBS case, and show that as $M$ increases, the regret increases. However, for large $T$, the effect of $M$ is minimal as the weights allocated to each sBS will go to zero. 
		\item We conduct extensive simulation results to corroborate our theoretical findings. In addition, we also show that the proposed \texttt{Bayesian estimation} method achieves a better performance compared to schemes such as Least Recently Frequently Used (LRFU), Least Frequently Used (LFU), and Least Recently Used (LRU) in terms of average cache hit. 
	\end{itemize}
	
	\textit{Notation:} Bold uppercase letter denotes matrices. $\mathbb{E}(\cdot)$ denotes the statistical expectation operator. $f(\cdot)$ represents the probability density function (PDF). Superscript $(\cdot)^{T}$ represents transposition. $||\cdot||_F$, $||\cdot||_{op}$ and $vec$ indicates the Frobenius norm, operator norm and vector respectively. $I_d$ represents the $d \times d$ identity matrix. Further, $\texttt{Dirch}(\alpha_1,\alpha_2,\ldots,\alpha_K)$ is the Dirichlet distribution with parameters $\alpha_1,\alpha_2,\ldots,\alpha_K.$

	\section{System Model and Problem Statement} \label{sec:sys_model}
\begin{figure}[hbtp]
		\centering
		\includegraphics[scale=0.35]{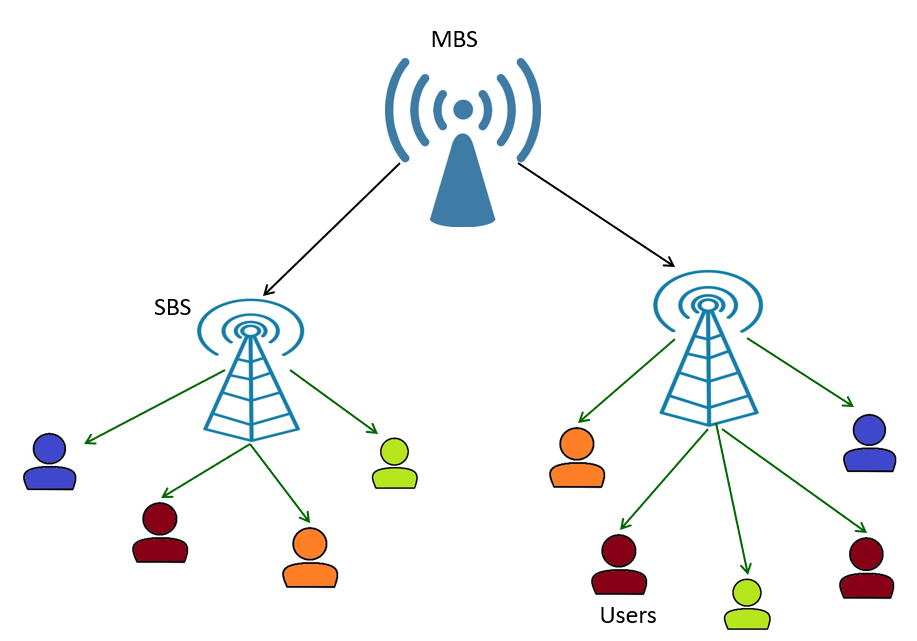}
		\caption{Distributed caching in a cellular network.}
		\label{fig:sysmodel}
	\end{figure}

	The system model consists of a wireless distributed content storage network with $M$ sBSs serving multiple users and one central MBS, as shown in Fig.~\ref{fig:sysmodel}. Each sBS can store up to $F$ contents/files of equal sizes from a catalog of contents denoted by $\mathcal{C}:=\{1,2,\ldots,F\}$. The requests are assumed to be independent and identically (iid) distributed across time.\footnote{A more general model of non-stationary requests can be handled based on the insights provided in the later part of our paper.} As we know, recommending a file influences the users request process, and hence recommendation can provide ``side information" about the future requests. In this paper, we consider the problem of jointly optimizing recommendation and caching policies in a cellular network. We model the influence of recommendation on the request via a conditional probability distribution denoted by $p_{ij,k}$, which represents the probability that a user requested a file $i$ from the sBS $k$ \emph{given} the content $j$ was recommended \cite{tony21}. Without loss of generality we assume that the time is slotted, and the PTM matrix for the $k$-th sBS denoted by $(\bm{P}_k)_{ij}:=p_{ij,k}$, $i,j = 1,2,\ldots,F$ is assumed to be fixed across time slots. For the sake of simplicity, it is assumed that at least one file is requested in every slot by each $N$th user in the network.\footnote{This can be ensured if the slot duration is chosen to be large enough.} Let us use $u_{i}$ and $v_{j}$ to represent the probabilities with which a file $i$ is cached, and the file $j$ is recommended at any sBS, respectively. This induces a set of caching and recommendation strategies denoted by
	\begin{equation} \label{eq:cache_const_set}
	\mathcal{C}_{c,r}:= \{(\bm{u},\bm{v}) \in [0,1]^{2 \times F}: \bm{u}^T \bm{1} \leq c, \bm{v}^T \bm{1} \leq r\},
	\end{equation}
	where $r$ and $c$ are recommendation and cache constraints, respectively. 
	In the sequel, the strategy is defined by the pair $(\bm{u},\bm{v})$. For a given strategy $(\bm{u},\bm{v}) \in \mathcal{C}_{c,r}$, the average cache hit at the sBS $k$ is given by $\bm{u}^T \textbf{P}_k\bm{v}$. If the matrix $\textbf{P}_k$ is known apriori at the sBS $k$, the optimal strategy can be found by solving $\max_{(\bm{u},\bm{v}) \in \mathcal{C}_{c,r}} \bm{u}^T \textbf{P}_k \bm{v}$.\footnote{Note that this optimization problem is bilinear, and hence in general hard to solve.} However, the matrix $\textbf{P}_k$ is unknown, and therefore it needs to be estimated from the demands. Let the variable $d_{k,i}^{(t)}$ denote the demand at the sBS $k$, and is defined as the total number of requests in the time slot $t$ for the file $i$. Since the demands arrive sequentially, the PTMs need to be estimated and updated in an online mode. The performance of such algorithms is measured in terms of regret. As opposed to the adversarial setting of online learning, here we have assumed that there is an underlying distribution from which the requests are generated, namely the PTM. Accordingly, the following provides the definition of the regret, which depends on the PTM.
	\begin{defn}(Regret)
		The regret at the sBS $k$ after $T$ time slots with respect to any sequence of strategies $(\bm{u}_{k,t},\bm{v}_{k,t})$, $t=1,2,\ldots,T$ is defined as 
		\begin{eqnarray} \label{eq:regret}
		\texttt{Reg}_{k,T} &: =& T \bm{u}_{k,*}^T \mathbf{P}_k \bm{v}_{k,*} - \sum_{t = 1}^{T} \bm{u}_t^T \mathbf{P}_k \bm{v_t},
		\end{eqnarray}
		where $(\bm{u}_{k,*}, \bm{v}_{k,*}) := \arg \max_{(\bm{u},\bm{v}) \in \mathcal{C}_{c,r}} \bm{u}^T \bm P_k \bm{v}$ is the optimal strategy at the sBS $k$.
	\end{defn}
	In this work, we provide answers to the following two questions: (i) how should one cache and recommend files in an online fashion that results in a sub-linear regret?, and (ii) how should a sBS use the caching and recommendation solutions of the neighboring sBSs to improve its own performance? Towards answering the first question, we propose two strategies at each SBS that result in a minimum regret. In particular, we consider two approaches that aim to find estimates of the PTMs in an online fashion, namely (i) \texttt{Point estimation} and (ii) \texttt{Bayesian estimation} methods, and solve the caching/recommendation problem. The first method is a naive method which acts as a benchmark while the second method balances the exploration and exploitation tradeoff that is typical in any regret minimization algorithm. Towards finding an answer to the second question above, we consider a linear combination of estimates of PTMs from the neighboring sBSs and find the coefficients that result in a smaller regret. In the following section, we provide caching and recommendation algorithms for single sBS scenario, and provide theoretical guarantees for them. In the later sections, we extend the analysis to multiple sBSs.
	\section{Joint Caching and Recommendation for Single sBS Scenario}	
	In this section, we consider a single sBS, i.e., $M = 1$. As mentioned above, using the demands obtained at the sBS, an estimate of the PTM matrix is computed using either \texttt{Point estimation} or \texttt{Bayesian estimation} method. Given an estimate $\hat{\bm{P}}^{(t)}_k$, the caching and recommendation strategies will be found by solving the following problem\footnote{For theoretical analysis, we assume that the problem can be solved exactly.}
	\begin{equation} \label{eq:dirch_cach_reco}
	(\hat{\bm{u}}_{o,t}^*, \hat{\bm{v}}_{o,t}^*) = \arg \max_{(\bm{u}, \bm{v}) \in \mathcal{C}_{c,r}} \bm{u}^T \hat{\bm{P}}^{(t)}_k \bm{v}.
	\end{equation}
	Now, we present the following two estimation procedures used in this paper. 	
	\begin{itemize}
		\item \texttt{Point estimation:} Given any SBS $k$, in this method, the demands until $t$ time slots is used to compute an estimate of the matrix $\mathbf{P}_k$. During the first $T$ time slots, recommendation and caching are done in an i.i.d. fashion with probabilities $q$ and $p$, respectively. Let $v_{jk}^{t} = 1$ if file $j$ was recommended in the slot $t-1$, and zero otherwise. The recommendation and caching constraints in \eqref{eq:cache_const_set} are satisfied by choosing $q:=r/F$ and $p:= c/F$. We can see that as the value of $t$ increases, the estimate becomes better, and hence results in a better performance. The estimate of the $ij$-th entry for the $k$th SBS of the $\mathbf{P}_k$ matrix is given by
		\begin{equation} \label{eq:point_est}
		\hat{p}_{ij,k}^{(t)} := \frac{\sum_{s=0}^{t-1} d_{ik}^{(s)} v_{jk}^{s-1} }{N \sum_{s=0}^{t-1} v_{jk}^{s-1}}.
		\end{equation}
		The above is a naive estimate of the probabilities by using a simple counting of events. The corresponding estimate of the matrix $\mathbf{P}_k$ be denoted by $\hat{\bm{P}_k}^{(t)}$. Since $\mathbb{E}\{\hat{p}_{ij,k}^{(t+1)} | \sum_{s=0}^{t-1} v_{jk}^s > 0\} = p_{ij,k}$, the point estimator is an unbiased estimator. Note that the regret obtained in the first $T$ slots will be maximum ($\mathcal{O}(T)$) as the caching and recommendations are done in a random fashion. However, to use this scheme as a benchmark, we assume an identical genie aided system where an estimate $\hat{\bm{P}_k}^{(t)}$ is available at time slot $t+1$ for caching and recommendation. Using this estimate, caching and recommendation are done by solving the optimization problem in \eqref{eq:dirch_cach_reco} with $\hat{\bm{P}}_k^{(t)}$ as the estimate in \eqref{eq:point_est} for all time slots $t=1,2,\ldots,T$. As we expect, the corresponding regret is small as the estimate at each time slot is good, and hence acts as a benchmark. 
		\item \texttt{Bayesian estimation:} In this method, for a given time slot, rows of the matrix $\bm{P}_k^{(t)}$ are sampled using a \textit{prior} distribution, which is updated based on the past demands. This may tradeoff the exploration versus exploitation while solving for the optimal recommendation and caching strategies. Here, Dirichlet distribution is chosen as a prior. The Dirichlet pdf is a multivariate generalization of the Beta distribution, and is given by
		\begin{equation} \label{eq:dirchlet}
		f(x_1,x_M, \alpha_1, \alpha_M) = \frac{\Gamma (\sum_{j = 1}^{M}\alpha_j)}{\prod_{j = 1}^{M}\Gamma(\alpha_j)} \prod_{j = 1}^{M}x_j^{\alpha_j - 1},
		\end{equation}
		$\alpha_j \geq 0$ $\forall$ $j$. The Dirichlet distribution is used as a conjugate pair in Bayesian analysis and the shape of the distribution is determined by the parameter $\alpha_j$. If $\alpha_j = 1$ $\forall$ $j$, then it leads to a uniform distribution. The higher the value of $\alpha_j$, the greater the probability of occurrence of $x_j$. The notation $(x_1,x_2,\ldots,x_M)  \sim \texttt{Dirch}(\alpha_1,\alpha_2,\ldots,\alpha_M)$ indicates that $(x_1,x_2,\ldots,x_M)$ is sampled from a Dirichlet distribution in \eqref{eq:dirchlet}. An estimate in the beginning of the time slot $t$ of the $i$-th row of the matrix $\hat{\bm{P}}_k^{(t)}$ is given by 
		\begin{eqnarray} \label{eq:bayes_est}
		(\hat{\bm{P}_k}^{(t)})_i \sim \texttt{Dirch}\Bigg(\sum_{s=1}^{t-1} d_{1k}^{(s)} v_{jk}^{s-1}, \sum_{i=1}^{t-1}d_{2k}^{(s)} v_{jk}^{s-1},  \sum_{s=1}^{t-1}d_{Fk}^{(s)} v_{jk}^{s-1}\Bigg), 
		\end{eqnarray}
		where $v_{jk}^{s-1}$ is as defined earlier with $v_{jk}^{0}$ sampled from $\{0,1\}$ with probability $q:=r/F$. 
		\begin{algorithm}[h] 
			\caption{Caching and recommendation algorithm (one sBS case) at any sBS $k$.} \label{alg:singleSBS}
			\begin{algorithmic}[1] 
				\Procedure{\texttt{Point estimation}/\texttt{Bayesian estimation}}{}
				\State $\hat{\bm{u}}_{b,0}^* \stackrel{i.i.d.}{\sim} \{0,1\}$ from $p = c/F$, \& $\hat{\bm{v}}_{b,0}^* \stackrel{i.i.d.}{\sim} \{0,1\}$ from $q = r/F$.
				\State \text{Recommend \& cache according to $\hat{\bm{v}}_{b,0}^*$ \& $\hat{\bm{u}}_{b,0}^*$.}
				\For{$t=0,1,\ldots,T$}
				\State \text{Observe demands $d^{(t)}_{ik}$ in slot $t$.}
				\State \text{Compute $\hat{\bm{P}}_{k}^{(t)}$ from \eqref{eq:point_est} for \texttt{Point }}
				\text{\texttt{estimation}} 
				\State \text{Compute $\hat{\bm{P}}_k^{(t)}$ from \eqref{eq:bayes_est} for \texttt{Bayesian}}
				\text{	 \texttt{estimation}}
				\State \text{Solve \eqref{eq:dirch_cach_reco}} // genie aided estimate for \texttt{point estimation} or \eqref{eq:bayes_est} for \texttt{Bayesian}
				\State \text{Use $(\hat{\bm{v}}_{b,t}^*, \hat{\bm{u}}_{b,t}^*)$ to recommend and cache.}				
				\EndFor
				\EndProcedure
			\end{algorithmic}
			
		\end{algorithm}
		After every time slot $t$, the recommendation and caching probabilities are selected by solving the optimization problem in \eqref{eq:dirch_cach_reco} with $\hat{\bm{P}}_k^{(t)}$ obtained in \eqref{eq:bayes_est}. A procedure to find a strategy is given in \textbf{Algorithm} $1$. 
	\end{itemize}
	In the following subsection, we provide theoretical guarantees of the above algorithm. 
	
	\subsection{Theoretical Guarantees} \label{sec:theoretical_guarant}
	In this section, we provide a high probability bound on the regret for both a genie aided \texttt{Point estimation} and \texttt{Bayesian estimation}. For the \texttt{Point estimation} case, we start by providing a lower bound on the waiting time which results in a performance that is $\epsilon$ close to the optimal performance. The result will be of the following form: With a probability of at least $1- \delta$, the following holds provided $t \geq \texttt{constant}$
	\begin{equation}
	\bm{u}_t^T \mathbf{P}_k \bm{v}_t \geq \sup_{(\bm{u},\bm{v}) \in \mathcal{C}_{c,r}} \bm{u}^T \mathbf{P}_k \bm{v} - \epsilon,
	\end{equation}
	where $(\bm{u}_t, \bm{v}_t)$ is the caching strategy obtained by using any algorithm. The constant $\epsilon$ depends on various parameters, as explained next. This result will be used to find a genie aided regret bound for the \texttt{point estimation} method. Towards stating theoretical guarantees, the following definition is useful. 
	\begin{defn}(Covering number) A set $\mathcal{N}_{\epsilon} := \{(\bm{x}_1,\bm{y}_1),(\bm{x}_2,\bm{y}_2),\ldots,(\bm{x}_{\mathcal{N}_{\epsilon}},\bm{y}_{\mathcal{N}_{\epsilon}})\}$ is said to be an $\epsilon$-cover of $\mathcal{C}_{c,r}$ if for any $(\bm{u},\bm{v}) \in \mathcal{C}_{c,r}$, there exists $(\bm{x}_j,\bm{y}_j) \in \mathcal{N}_{\epsilon}$ for some $j$ such that $\norm{\bm{u} - \bm{x}_j} \leq \frac{\epsilon}{8}$ and  $\norm{\bm{v} - \bm{y}_j} \leq \frac{\epsilon}{8}$.
	\end{defn}
	
	The following theorem provides a bound that is useful to provide the final result.
	
	\begin{theorem} \label{thm:bound_first_general}
		For a given estimate of the PTM denoted by $\hat{\bm{P}}_k^{(t)}$ using \texttt{Point estimation} or \texttt{Bayesian estimation}, the following holds good
		\begin{eqnarray}
		\Pr\left\{\sup_{(\bm{u},\bm{v}) \in \mathcal{C}_{c,r}} \bm{u}^T \mathbf{P}_k \bm{v} - \bm{u}_t \mathbf{P}_k\bm{v}_t \geq \epsilon\right\}  \leq |\mathcal{N}_{\epsilon}| \Pr\left\{\norm{\widehat{\Delta P}^{(t)}}_F \geq \frac{\epsilon}{4 \kappa r c}\right\}, \label{eq:thm_point_est_bound}
		\end{eqnarray}
		where $(\bm{u}_t, \bm{v}_t)$ is the output of the Algorithm \ref{alg:singleSBS} at time $t$, and $\widehat{\Delta P}^{(t)}:= \mathbf{P}_k - \hat{\bm{P}}_k^{(t)}$. Further, $\kappa > 0$ is some constant. 
	\end{theorem}
	\emph{Proof:}  See Appendix \ref{appendix:a}.
	
	Using the above result, in the following, we provide our first main result on the performance of the \texttt{Point estimation} scheme. 
	\begin{theorem} \label{thm:point_est_guarant_1bs}
		Using \eqref{eq:dirch_cach_reco} for caching and recommendation in slot $t$, for any $\epsilon > 0$, with a probability of at least $1-\delta$, $\delta > 0$, $(\bm{u}_{o,t}^*)^T \mathbf{P}_k \bm{v}_{o,t}^* \geq \sup_{(\bm{u},\bm{v}) \in \mathcal{C}_{c,r}} \bm{u}^T \mathbf{P}_k \bm{v} - \epsilon$, provided
		\vspace{0.5cm}
		\begin{equation}  \label{eq:bound_ontime_single_BS}
		t \geq \frac{1}{q \left(1 - \exp\{-\frac{N\epsilon^2}{8 \kappa^2 F^2 c^2 r^2}\}\right)} \log \frac{2 \mathcal{N}_\epsilon F^2}{\delta}.
		\end{equation}
	\end{theorem}
	\emph{Proof:}  See Appendix \ref{appendix:b}.
	
	As we know, the regret achieved by the \texttt{Point estimation} method is $\mathcal{O}(T)$ as it incurs non-zero constant average error for all the slots $t$ satisfying \eqref{eq:bound_ontime_single_BS}. In this method, the estimation of PTM is done using the samples obtained from the first $t$ slots, and the caching strategy is decided based on this estimate. However, an improvement over this is to continuously update the estimates, and the caching/recommendation strategies. Instead of analyzing the regret for this, we assume that at any time slot $t$, a genie provides an estimate of the PTM as in \eqref{eq:point_est} to compute the caching/recommendation strategies, and provide the corresponding approximate regret bound. In particular, in Appendix \ref{sec:theoretical_guarant_point}, we show the following bound on the regret for a genie aided \texttt{point estimation} method. 
	\begin{theorem}
		With a probability of at least $1-1/T$, a regret of $\mathcal{O}(T^{2/3}\sqrt{\log T})$ can be achieved through the genie aided \texttt{Point estimation} method.   
	\end{theorem}
	It can be observed that the regret scales faster than $\sqrt{T}$. In the following, we present the result for \texttt{Bayesian estimation} method, and contrast the result with the genie aided case. 
	
	\subsection{Bayesian Estimation: Single sBS Scenario} \label{sec:theoretical_guar_bayes_sing_BS}
	Note that unlike the analysis for \texttt{Point estimation}, in this case, the strategies are correlated across time, which makes the analysis non-trivial. The approach we take is to convert a sequence of random variables (function of caching and recommendation across time) into a Martingale difference. This enables us to use the Azuma's inequality, which can be used to provide high probability result on the regret. In the following, we provide the result. 
	\begin{theorem}  \label{thm:bay_guarant_1bs}
		For the \texttt{Bayesian estimation} in Algorithm \ref{alg:singleSBS}, for any $\epsilon > 0$, with a probability of at least $1-1/T$, the following bound on the regret holds
		\begin{eqnarray}
		\texttt{Reg}_T \leq {2rc\max_{ij} p_{ij} |\mathcal{N}_\epsilon|}\sum_t\exp\bigg\{\frac{-8\psi_t^2}{cr\abs{\mathcal{N}_\epsilon}^2 \bar{\sigma}_t^2(t)}  \bigg\} + {2\sum_t\psi_t} + \sqrt{128r^2c^2T\log(T)},
		\end{eqnarray}
		where $\alpha_{ij}^{(t)} = \sum_{q=1}^{t-1}d_{i}^{(q)}v_{j}^{(q-1)}$, $\bar{\sigma}_t^2:= \left[\sum_{j=1}^F \frac{1}{\left(\sum_i \alpha_{ij}^{(t)} +1\right)^2}\right]$, and $\psi_t > 0$.
	\end{theorem} 
	\begin{figure}[hbtp]
		\centering
		\includegraphics[scale=0.4]{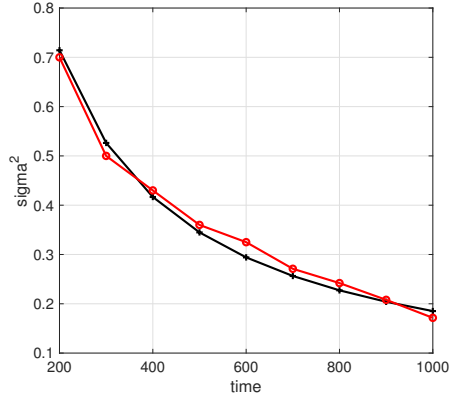}
		\caption{Plot of $\bar{\sigma}_t^2$ versus time. The plot also shows that $\mathcal{O}(1/t)$ is a good fit for $\bar{\sigma}_t^2$.}
		\label{fig:sigmaplot}
	\end{figure}
	\emph{Proof:}  See Appendix \ref{app:regret_oneBS}.\\
	\\
	\textbf{Remark:} Note that the above result is an algorithm and data dependent bound as it depends on the recommendation strategy and the demands. As a consequence, the choice of $\psi_t$ to obtain better regret is not clear. In order to provide more insights into the result, we plot $\bar \sigma_t^2$ versus time slot in Fig.~\ref{fig:sigmaplot}. In the same plot, we have also shown that $\mathcal{O}(1/t)$ is a good fit for $\bar \sigma_t^2$. Furthermore, the cardinality of the $\epsilon$ cover does not scale with $T$. Thus, by choosing $\psi_t = 1/t^a$, the regret becomes $\mathcal{O}(T^a) + \mathcal{O}(T^{1-a}) + T^{1/2}$, where $a \leq 1/2$. Thus, by choosing $a = 1/2$ results in a $\sqrt{T}$ regret. Recall that an approximate regret of $\mathcal{O}(T^{2/3} \sqrt{\log T})$ is shown for the genie aided case while the \texttt{Bayesian estimation} method achieves a regret of the order $\sqrt{T}$. In other words, the Bayesian performance is better than the genie aided regret in the point estimation case by a factor of $T^{1/6}$. In the next section, we extend our results to two sBS scenario. 
	\section{Proposed Caching and Recommendation Strategies With Multiple sBSs}
	In this section, we present caching and recommendation algorithms when there are multiple sBSs. In particular, we provide insights on how to use the neighboring sBSs estimates to further improve the overall caching and recommendation performance of the network. First, we present the results for two sBS scenario, and similar analysis will be used to extend the results to multiple sBSs. 
	\subsection{Two Small Base Station Scenario}
	In this subsection, we consider a two sBSs scenario connected with the same MBS. As described in Section~\ref{sec:sys_model}, $\textbf{P}_1$ and $\textbf{P}_2$ represent PTMs for sBS-1 and sBS-2, respectively. The central MBS sends the global update of the recommendation and caching decisions to each sBS. Assume that the request across sBSs are independent. Let each sBS use one of the estimation methods in Algorithm \ref{alg:singleSBS}. Let $\hat{\textbf{P}}_1^{(t)}$ and $ \hat{\textbf{P}}_2^{(t)}$ be the corresponding estimates (either \texttt{point} or \texttt{Bayesian estimate}) of $\textbf{P}_1$ and $\textbf{P}_2$, respectively. The two sBSs convey their respective PTM to the central MBS. The central MBS computes an estimate $\hat{\textbf{Q}}_k^{(t)}$, $k=1,2$ for sBS $1$ and sBS $2$ as a linear combination of the two estimates as given below
	\begin{equation} \label{eq:two_bs_est}
	\hat{\textbf{Q}}_k^{(t)} = \lambda_k \hat{\textbf{P}}_1^{(t)} + (1 - \lambda_k) \hat{\textbf{P}}_2^{(t)}, 
	\end{equation}
	where $\lambda_k \in [0,1]$, $k=1,2$ strikes a balance between the two estimates. The above estimate is used to compute the respective caching and recommendation strategies for the two sBSs and will be communicated to the respective sBSs. The above results in a better estimate, for example, when $\textbf{P}_1 = \textbf{P}_2$ or when the two matrices are close to each other. The corresponding algorithm is shown below. First, we prove the following guarantee for the \texttt{Point estimation} method. 
	
	\begin{theorem}  \label{thm:point_est_guarant_2bs}
		For Algorithm \ref{alg:twoSBS} with \texttt{Point estimation}, for any sBS $k$ and for any $\epsilon > 0,$ 
		with a probability of at least $1-\delta$, $\delta > 0$, the regret $\texttt{Reg}_{k,T} < \epsilon$, i.e., $\Pr\bigg\{(\bm{u}_{k,t}^*)^T \textbf{P}_k \bm{v}_{k,t}^* \geq \sup_{(\bm{u},\bm{v}) \in \mathcal{C}_{c,r}} \bm{u}^T \hat{\bm{Q}}^{(t)}_k \bm{v} - \epsilon_k \bigg\} > 1- \delta$ provided 
		\begin{equation} \label{eq:t_bound}
		t \geq \max\left\{\tau\left(\frac{\epsilon_k}{\lambda_k}, \frac{\delta}{2}\right), \tau\left(\frac{\epsilon_k}{(1-\lambda_k)}, \frac{\delta}{2}\right)\right\}, 
		\end{equation}
		where 
		\begin{equation}
		\tau(\epsilon, \delta) := \frac{1}{q \left(1 - \exp\{-\frac{N\epsilon^2}{8 \kappa^2 F^2 c^2 r^2}\}\right)} \log \frac{2 |\mathcal{N}_\epsilon| F^2}{\delta}.
		\end{equation}
		Further, $\epsilon_k: = \epsilon/2 - (1 - \lambda_k) \sup_{(\bm{u},\bm{v}) \in \mathcal{C}_{c,r}} \abs{\bm{u}^T (\textbf{P}_2 - \textbf{P}_1) \bm{v}}$.
	\end{theorem} 
	\emph{Proof:}  See Appendix \ref{appendix:d}.
	
	\begin{algorithm}[h] 
		\caption{Caching and recommendation algorithm (two sBS case)} \label{alg:twoSBS}
		\begin{algorithmic}[1] 
			\Procedure{\texttt{Point estimation}/\texttt{Bayesian estimation}}{}
			\State $\hat{\bm{u}}_{b,0}^* \stackrel{i.i.d.}{\sim} \{0,1\}$ from $p = c/F$, and $\hat{\bm{v}}_{b,0}^* \stackrel{i.i.d.}{\sim} \{0,1\}$ from $q = r/F$.
			\State \text{Recommend \& cache according to $\hat{\bm{v}}_{b,0}^*$ \& $\hat{\bm{u}}_{b,0}^*$.}
			\For{$t=0,1,\ldots,T$}
			\State \text{Observe demands $d^{(t)}_{ijk}$ in slot $t$, $k=1,2$.}
			\If{\texttt{point estimation}}
			\State \text{Compute $\hat{\bm{P}}_{k}^{(t)}$ from \eqref{eq:point_est} for \texttt{point }}
			\text{\texttt{estimation}} 
			\Else
			\State \text{Compute $\hat{\bm{P}}_k^{(t)}$ from \eqref{eq:bayes_est} for \texttt{Bayesian }}
			\texttt{\texttt{estimation}}
			\EndIf
			\State \text{Choose $\lambda_i$, and find $\hat{\bm{Q}}^{(t)}_i$ from \eqref{eq:two_bs_est}, and }
			\text{solve $	(\hat{\bm{u}}_{k,t}^*, \hat{\bm{v}}_{k,t}^*) = \arg \max_{(\bm{u}, \bm{v}) \in \mathcal{C}_{c,r}} \bm{u}^T \hat{\bm{Q}}^{(t)}_k \bm{v}$}
			\State \text{Use $(\hat{\bm{v}}_{k,t}^*, \hat{\bm{u}}_{k,t}^*)$ to recommend and cache.}		
			\EndFor
			\EndProcedure
		\end{algorithmic}
	\end{algorithm}
	As in the single sBS case, to benchmark the performance of \texttt{Bayesian estimation} method, we consider a genie aided scenario, and in Appendix~\ref{app:2sbs_heuristics}, we show that it achieves an approximate regret of 
	\begin{eqnarray}
	\texttt{Reg}_{k,T} \lessapprox \max \Bigg\{ \Theta \lambda_k  , \Theta(1-\lambda_k)  \Bigg\}  T^{2/3} + 2T(1 - \lambda_k)\mathcal{V}_{12}, \nonumber
	\end{eqnarray}
	where $\mathcal{V}_{12}: = \sup_{(\bm{u},\bm{v}) \in \mathcal{C}_{c,r}} \abs{\bm{u}^T (\textbf{P}_2 - \textbf{P}_1) \bm{v}}$ and $\Theta = \sqrt[3]{\frac{8 \kappa^2 F^2 c^2 r^2(\log 4F^2T^2 + F)}{q N}} $. Note that when the second term is non-zero, i.e., $\bm P_1 \neq \bm P_2$, the above clearly shows the trade-off between the two terms. The first term scales as $T^{2/3}$ while the second term scales with $T$ linearly. This can be balanced by using $\lambda_k = 1 - \frac{1}{\sqrt{T}}$, which results in  $\mathcal{O}(\sqrt T)$ scaling of regret. Note that the choice $\lambda_k = 1 - \frac{1}{\sqrt{T}}$ reveals that as time progresses, i.e., as the sBS $k$ collects more samples, the weights allocated to the neighboring sBS should go down to zero, as expected. Furthermore, by appropriately choosing $\lambda_k$ as above, the regret obtained is of the order $T^{2/3}$. On the other extreme when $\bm P_1 = \bm P_2$, the second term is zero. In this case, the optimal choice is $\lambda_k = 1/2$, as expected. Next, we present the guarantees for Algorithm \ref{alg:twoSBS}.
	
	\begin{theorem}  \label{thm:bay_est_guarant_2bs}
		For Algorithm \ref{alg:twoSBS} with \texttt{Bayesian estimation}, for 
		\begin{equation} \label{eq:epsilon_bound_multi_bs}
		\epsilon > 2 \max_{k=1,2} (1 - \lambda_k) T \sup_{(\bm{u},\bm{v}) \in \mathcal{C}_{c,r}} \abs{\bm{u}^T (\textbf{P}_1 - \textbf{P}_2) \bm{v}},
		\end{equation}
		with probability of at least $1-\delta$, $\delta > 0$, for any BS $k \in \{1, 2\}$, the regret can be bounded as	
		\begin{equation} 
		\texttt{Reg}_{k,T}  \leq \max\left\{R_k\left(\frac{2\epsilon_k}{\lambda_k}, \frac{\delta}{2}\right), R_k\left(\frac{2\epsilon_k}{(1-\lambda_k)}, \frac{\delta}{2}\right)\right\}. 
		\end{equation}
		In the above,
		\begin{eqnarray} \label{eq:regret_multi_sbs}
		R_k(\epsilon,\delta) := 2rc\max_{ijk} p_{ijk} |\mathcal{N}_\epsilon|\sum_t\exp\bigg\{\frac{-8\psi_t^2}{cr\abs{\mathcal{N_\epsilon}}^2 \bar{\sigma}^2_k(t)}  \bigg\} + 2\sum_t\psi_t + \sqrt{128r^2c^2T\log(1/\delta)},
		\end{eqnarray}
		$\alpha_{ijk}^{(t)} = \sum_{q=1}^{t-1}d_{ik}^{(q)}v_{jk}^{(q-1)}$, $\epsilon_k: = \epsilon/2 - (1 - \lambda_k) T \sup_{(\bm{u},\bm{v}) \in \mathcal{C}_{c,r}} \abs{\bm{u}^T (\textbf{P}_1 - \textbf{P}_2) \bm{v}}$, and $\bar{\sigma}^2_k(t):= \left[\sum_{j=1}^F \frac{1}{\left(\sum_i \alpha_{ijk}^{(t)} +1\right)^2}\right]$.
	\end{theorem}
	\emph{Proof:}  See Appendix \ref{appendix:e}.\\
	\\
	\textbf{Remark:} The result shows the trade-off exhibited by $\lambda_k$. In particular, larger $\lambda_k$ makes the first regret inside the $\max$ term in \eqref{eq:regret_multi_sbs} larger, and smaller $\lambda_k$ ensures that the second term inside the $\max$ above dominates. Similar to the single sBS scenario, using $\lambda_k = 1 - \frac{1}{\sqrt{T}}$, results in $\mathcal{O}(\sqrt T)$ scaling of regret. Further, the above result is an algorithm dependent bound as the bound depends on the recommendation strategy, which is determined by the algorithm. Following the single sBS case, we can show similar regret of $\mathcal{O}(\sqrt{T})$, which is superior to the \texttt{point estimation} method. In the simulation results section, we present more details on this trade-off in the finite $T$ regime. In the next section, we extend the analysis to multiple sBSs. 
	
	\subsection{Multiple Small Base Station Scenario}
	In this section, we extend the analysis and algorithm of the previous section to heterogeneous network with $M$ sBSs connected to a central MBS. The requests at each sBS are assumed to be i.i.d. with PTM $\textbf{P}_1, \textbf{P}_2, \ldots, \textbf{P}_M$ as described in Section~\ref{sec:sys_model}. Similar to the two sBS model, each sBS computes an estimate of the PTM as follows
	\begin{equation} \label{eq:mult_bs_est}
	\hat{\textbf{Q}}_k^{(t)} = \lambda^{(k)}_1 \hat{\textbf{P}}_1^{(t)} + \lambda_2^{(k)} \hat{\textbf{P}}_2^{(t)} + \ldots + \lambda^{(k)}_M \hat{\textbf{P}}_M^{(t)},
	\end{equation}
	where $\lambda^{(k)}_1, \lambda^{(k)}_2, \ldots, \lambda^{(k)}_M$, $k = 1,2,\ldots,M$ are non-zero coefficients to be determined later that satisfy $\sum_{j=1}^M \lambda_j^{(k)} = 1$. The following theorem is a generalization of two BS model which provides a guarantee on the minimum time required to achieve a certain level of accuracy with high probability.
	\begin{theorem} \label{thm:point_est_guarant_mbs}
		Using \eqref{eq:mult_bs_est} for \texttt{point estimation}, for any $$\epsilon > M^2 \max_{k} \{ (1 - \lambda_1^{(k)}) \mathcal{D}_{1} - \lambda_2^{(k)} \mathcal{D}_2 - \ldots  - \lambda_M^{(k)} \mathcal{D}_M\},$$ with a probability of at least $1-\delta$, $\delta > 0$, for any BS $k$, the regret $\texttt{Reg}_{k, T} < \epsilon$ 
		provided 
		\begin{eqnarray} 
		t &\geq& \max_{j =1,2,\ldots,M}\Bigg\{\tau\left(\frac{\epsilon_j}{\lambda_j^{(k)}}, \frac{\delta}{M}\right) \Bigg\}, \label{eq:t_bound_msbs}
		\end{eqnarray}
		where 
		\begin{equation}	
		\tau(\epsilon, \delta) :=  \frac{1}{q \left(1 - \exp\{-\frac{N\epsilon^2}{8 \kappa^2 F^2 c^2 r^2}\}\right)} \log \frac{2 \mathcal{N}_\epsilon F^2}{\delta},		
		\end{equation}
		$\epsilon_{k} := \epsilon/M^2 - (1 - \lambda_1^{(k)}) \mathcal{D}_{1} + \lambda_2^{(k)} \mathcal{D}_2 +, \ldots,  + \lambda_M^{(k)} \mathcal{D}_M$, and $\mathcal{D}_{k}: = \sup_{(\bm{u},\bm{v}) \in \mathcal{C}_{c,r}} \abs{\bm{u}^T P_k \bm{v}}$ $\forall$ $k$ $= {1,2,\ldots, M}$.
	\end{theorem}
	\emph{Proof:}  See Appendix \ref{appendix:f}.
	
	In Appendix \ref{app:regret_multisbs_heuristics_genieaided}, we show that the regret for the genie aided case after appropriate choice for $\lambda_k$ is given by
	\begin{eqnarray}
	\texttt{Reg}_{k,T} \lessapprox \Theta M^2 \max \Bigg\{  \lambda_1^{(k)}  , \ldots, \lambda_M^{(k)}  \Bigg\} T^{2/3} + M^2T\big\{(1 - \lambda_1^{(k)}) \mathcal{D}_{1} - \ldots  -\lambda_M^{(k)} \mathcal{D}_M\big\} \nonumber 
	\end{eqnarray}
	where $\mathcal{D}_{k}: = \sup_{(\bm{u},\bm{v}) \in \mathcal{C}_{c,r}} \abs{\bm{u}^T P_k \bm{v}}$ $\forall$ $k$ $= {1,2,\ldots, M}$ and $\Theta = \sqrt[3]{\frac{8 \kappa^2 F^2 c^2 r^2(\log F^2T^2 + F)}{q N}  }$. 
	\\
	
	\textbf{Remark:} Note that the value of regret depends on the values of $\lambda_1^{(k)}, \ldots, \lambda_M^{(k)}$ and the term $\mathcal{D}_k$. The first term scales as $T^{2/3}$ while the second term scales with $T$ linearly. Using $\lambda_i^{(k)} = 1 - \frac{1}{(M-1)\sqrt{T}}$ for $i=1,2,\ldots,M$ results in a balance between the two terms. In particular, this leads to a regret that scales as $\mathcal{O}(T^{2/3})$. Similar to the single sBS case, the choice $\lambda_M^{(k)} = \frac{1}{(M-1)\sqrt{T}}$ reveals that as time progresses, i.e., as the sBS $k$ collects more samples, the weights allocated to the neighboring sBS should go down to zero, as expected. For finite $T$, one can optimize the above regret with respect to $\lambda_M^{(k)}$'s, and find the optimal choice; this is relegated to our future work. Next we present the regret bound for the \texttt{Bayesian estimation} method. 
	
	\begin{theorem} \label{thm:bays_est_guarant_mbs}
		Using \eqref{eq:mult_bs_est} for \texttt{Bayesian estimation}, for any 
		$$\epsilon> M^2 \max_{k}  \{ (1 - \lambda_1^{(k)}) \mathcal{I}_{1} - \lambda_2^{(k)} \mathcal{I}_2 -, \ldots,  - \lambda_M^{(k)} \mathcal{I}_M\},$$ 
		with probability of at least $1-\delta$, $\delta > 0$, for any BS $k \in \{1,2, \ldots, M\}$, we have the following bound on the regret
		\begin{eqnarray} 
		\texttt{Reg}_{k,T}  &\leq& \max_{j=1,2,\ldots,M} R_k\left(\frac{\epsilon_j}{\lambda_j^{(k)}},\frac{\delta}{M}\right), 
		\end{eqnarray}
		where $R_k(\epsilon, \delta)$ is as defined in Theorem \ref{thm:bay_guarant_1bs} for the single sBS, and 
		$\epsilon_{k} := \epsilon/M^2 - (1 - \lambda_1^{(k)}) \mathcal{I}_{1} + \lambda_2^{(k)} \mathcal{I}_2 +, \ldots,  + \lambda_M^{(k)} \mathcal{I}_M$ with  
		$\mathcal{I}_{k}: = T\sup_{(\bm{u},\bm{v}) \in \mathcal{C}_{c,r}} \abs{\bm{u}^T P_k \bm{v}}$. 
	\end{theorem}
	\emph{Proof:}  See Appendix \ref{appendix:g}.\\ 
	\\
	\textbf{Remark:} As in the case of single and two sBS scenarios, the above result is an algorithm dependent bound as it depends on the recommendation strategy. Using $\lambda_1^{(k)} = 1 - \frac{1}{(M-1)\sqrt{T}}$ and $\lambda_M^{(k)} = \frac{1}{(M-1)\sqrt{T}}$, results in $\mathcal{O}(\sqrt T)$ scaling of regret. Following the single sBS analysis, one can show that even in the multiple sBS scenario also, the regret is superior to the \texttt{point estimation} method and scales as $\mathcal{O}(\sqrt{T})$. Clearly, the regret obtained is better than the genie aided scenario whose regret scales as $\mathcal{O}(T^{2/3})$. In the next section, we present experimental results that corroborates our theoretical observations.

\section{Simulation Results}
In this section, simulation results are presented to highlight performance of the proposed caching and recommendation model. The simulation setup consists of multiple sBSs with multiple users. We assume a time-slotted system in the simulation setup. For the heterogeneous model, the simulation consists of two scenarios as follows:
\begin{itemize}
	\item Fixed Link Scenario: In this case, the links between sBS and users are uniformly and independently distributed in $\{0, 1\}$ with probability $1/2$. 
	\item \texttt{SINR} based Scenario: In this case, the sBS and users are assumed to be distributed uniformly in a geographical area of radius $500$m. It is assumed that a sBS and users can communicate only if the corresponding \texttt{SINR} is greater than a threshold. This \texttt{SINR} takes into account the fading channel, the path loss, power used, and the distance between the user and the sBS.
	The minimum rate at which a file can be transferred from the sBS to a user is given by the threshold, and hence the reciprocal of the rate indicates the delay. In the simulation, we have used $\tau := \frac{1}{\log(1+\texttt{SINR})}$ as a measure of the delay between a user and a sBS. However, when the requested file is absent, a backhaul fetching delay of $\alpha \times \tau$ is counted in addition to the downlink delay of $\tau$, i.e., the overall delay when the file is absent is $(\alpha + 1) \tau$, with $\alpha = 10$. Also, if the threshold is $R$, then at least $R$ bits can be sent in a time duration of at most $1/\log(1 + \texttt{SINR})$ seconds, and hence the throughput is roughly $R\log(1 + \texttt{SINR})$ bits/second.
\end{itemize}

\begin{figure}
	\centering
	\begin{minipage}{.5\textwidth}
		\centering
		\includegraphics[scale = 0.3]{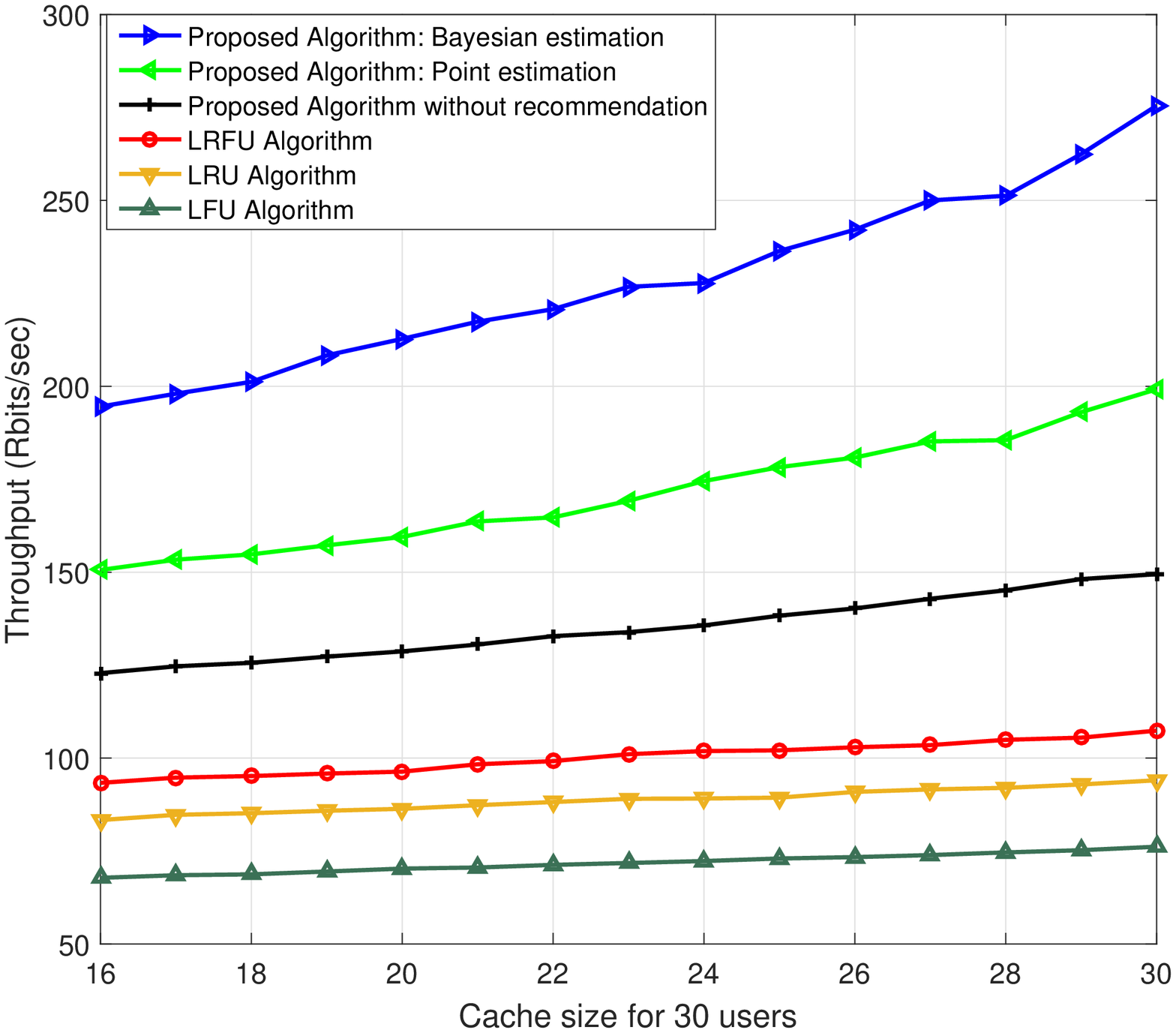}
		\caption{ \scriptsize{Average throughput v/s cache size for $15$ sBS model.}}
		\label{fig:comp_hetero}
	\end{minipage}%
	\begin{minipage}{.5\textwidth}
		\centering
	\includegraphics[scale = 0.28]{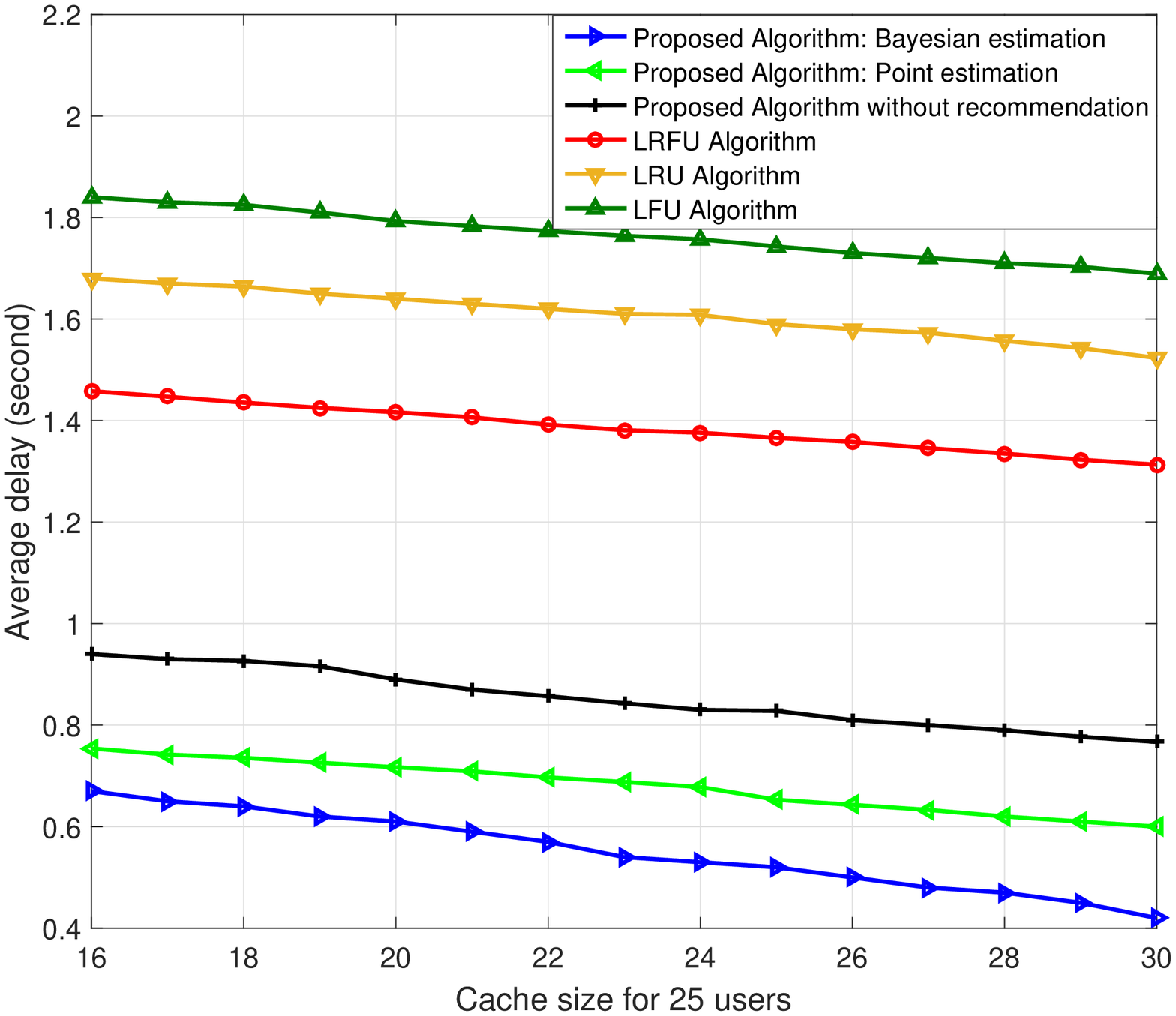}
	\caption{\scriptsize{Average delay v/s cache size for 2-sBS model.}}
	\label{fig:comp2sbs}
	\end{minipage}
\end{figure}

Fig.~\ref{fig:comp_hetero} shows the throughput plot for the considered heterogeneous system with $5$ sBS and $30$ users. In Fig.~\ref{fig:comp_hetero}, the total number of files and threshold value for \texttt{SINR} are $100$, and $12$, respectively. The throughput for the proposed algorithm with recommendation is $225$ bits/s for a cache size of $24$, while LRFU, LRU and LFU algorithm have a throughput of $100$ bits/s, $85$ bits/s and $70$ bits/s respectively for the same cache size. Thus, from Fig.~\ref{fig:comp_hetero} we can see that the proposed algorithm has  higher throughput as compared to the existing algorithms.

Fig.~\ref{fig:comp2sbs} corresponds to the \texttt{SINR} scenario for a two sBS model. Fig.~\ref{fig:comp2sbs} shows the average delay versus cache size plot for (a) cache placement algorithm with recommendation, (b) cache placement algorithm without recommendation, (c) LRFU algorithm, (d) LRU algorithm, and (e) LFU algorithm. In Fig.~\ref{fig:comp2sbs}, the number of sBSs, the number of users, the total number of files, and the threshold value for \texttt{SINR} are $2$, $25$, $100$ and $12$dB, respectively. From Fig.~\ref{fig:comp2sbs} we can observe that the delay of both the proposed algorithms is less as compared to the other benchmark algorithms, since pre-fetching files according to the estimated methods results in lower fetching costs from the backhaul and hence less delay. 

\begin{figure}
	\centering
	\begin{minipage}{.5\textwidth}
		\centering
		\includegraphics[scale = 0.4]{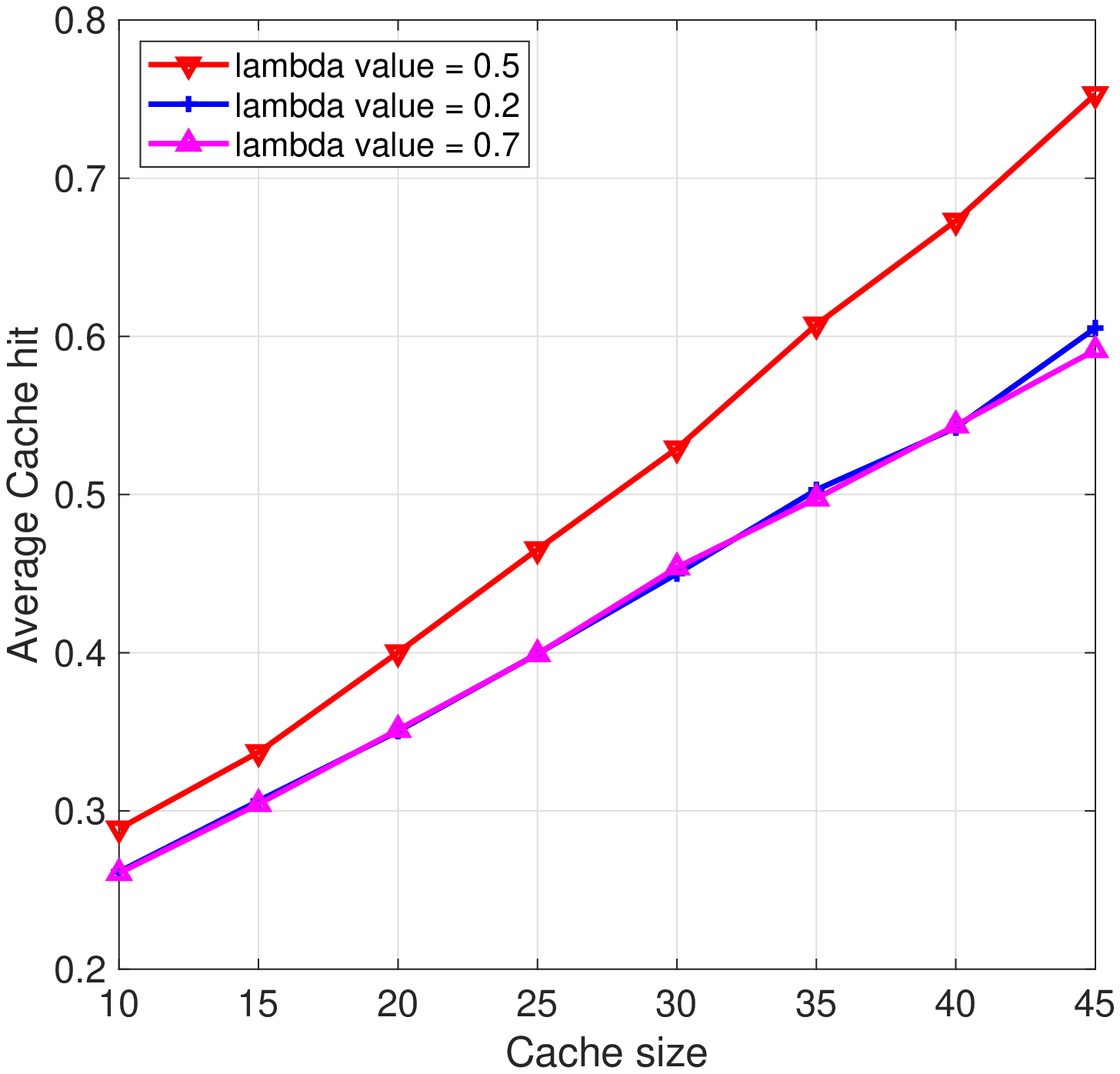}
	\caption{\scriptsize{Cache hit v/s cache size for 2 sBS and $\textbf{P}_1 = \textbf{P}_2$.}}
	\label{fig:lambda}
	
	\end{minipage}%
	\begin{minipage}{.5\textwidth}
		\centering
		\includegraphics[scale = 0.38]{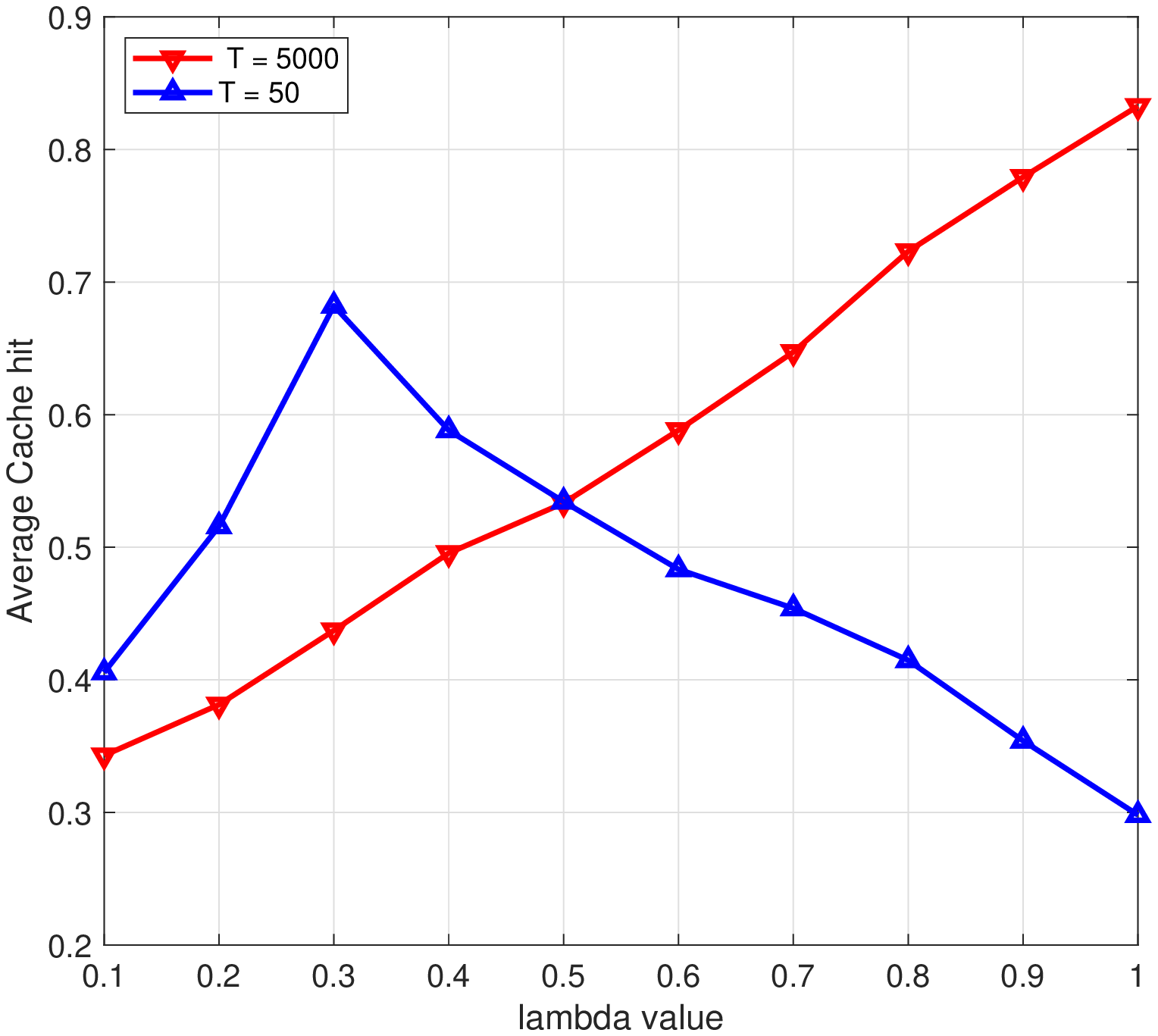}
	\caption{\scriptsize{Cache hit v/s $\lambda$ for 2 sBS and $\textbf{P}_1 \neq \textbf{P}_2$.}}
	\label{fig:lambda_diff}	
	
	\end{minipage}
\end{figure}

Fig.~\ref{fig:lambda} shows the plot for two sBS model. The value of $\lambda_1$ is varied between 0.1 and 1. From the Fig.~\ref{fig:lambda}, we can observe that as the value of $\lambda_1$ approaches 0.5, the average cache hit increases. This is because for $\lambda_1 = 0.5$ and $P_1 = P_2$ , the $Q$ popularity profile matrix of MBS will have maximum similarity to the individual sBS popularity profile matrix and hence the cache hit will be maximum for $\lambda_1 = 0.5$ and it will gradually decrease as we further increase the value of $\lambda_1$. 
Fig.~\ref{fig:lambda_diff} shows the plot for  average cache hit versus $\lambda$ for $2$ sBS when $\textbf{P}_1 \neq \textbf{P}_2$. From the Fig.~\ref{fig:lambda_diff}, we can observe that for larger $T$, the optimal lambda value is close to $1$. Also, for smaller value of $T$, depending on the value of $\Theta$, the optimal value of $\lambda$ is less than $1$ and as shown in Fig.~\ref{fig:lambda_diff}, the optimal value of $\lambda$ is  $0.3$.
Thus, the simulation results prove that the recommendation helps in increasing the average cache hit when compared to the algorithm without recommendation and it also performs better than the existing popular LRFU, LRU and LFU algorithms.

\section{Conclusion}
In this paper, we have proposed a novel joint caching decision along with recommendation in the upcoming next generation cellular networks. We leverage the implications of recommendation on user requests and the overall average cache hit is improved. Two estimation methods, \texttt{Bayesian estimation} and \texttt{Point estimation} are used to determine the user request pattern. An algorithm is then proposed to jointly optimize caching and recommendation. A multi-tier heterogeneous model consisting of MBS and sBSs is also presented and an approximately high probability bound on the regret for both the estimation method is provided. Finally, simulation results and theoretical proofs support the superior performance of the proposed method over the existing algorithms.

\bibliographystyle{IEEEtran}
\bibliography{strings}

\appendices
\section{Proof of Theorem \ref{thm:bound_first_general}} \label{appendix:a}
From \cite{bousquet2003introduction}, it follows that  

\begin{eqnarray} \label{eq:first_eq}
\sup_{(\bm{u},\bm{v}) \in \mathcal{C}_{c,r}} \hspace{-0.3cm}\bm{u}^* \bm{P}_k \bm{v}^* - \bm{u}^T \bm{P}_k \bm{v} \leq 2 \sup_{(\bm{u},\bm{v}) \in \mathcal{C}_{c,r}}\hspace{-0.3cm} \abs{\bm{u}^T \widehat{\Delta P}^{(t)} \bm{v}}.
\end{eqnarray}
Let $\bm{x}^{*}$ and $\bm{y}^*$ be solutions to $\sup_{(\bm{u},\bm{v}) \in \mathcal{C}_{c,r}} \abs{\bm{u}^T \widehat{\Delta P}^{(t)} \bm{v}}$. Since $\bm{x}^{*}$ and $\bm{y}^*$ belong to $\mathcal{C}_{c,r}$, for some $i=1,2,\ldots, \mathcal{N}_\epsilon$, there exist $\bm{x}_i$ and $\bm{y}_i$ in $\mathcal{A}_{\epsilon}$ such that $\norm{\bm{x}^{*} - \bm{x}_i}_2 \leq \epsilon/8$, and $\norm{\bm{y}^{*} - \bm{y}_i}_2 \leq \epsilon/8$. Further, by adding and subtracting $\bm{x}_i  \widehat{\Delta P}^{(t)} \bm{y}_i$ and $\bm{x}_i  \widehat{\Delta P}^{(t)} \bm{y}_i$, we get 
\begin{equation} \label{eq:second_eq}
\abs{({\bm{u}^*})^T \widehat{\Delta P}^{(t)} \bm{v}^*} \leq \abs{\bm{x}_i \widehat{\Delta P}^{(t)} \bm{y}_i} + \frac{\epsilon \norm{\widehat{\Delta P}^{(t)}}_{op}} {4}.
\end{equation}
From \eqref{eq:first_eq} and \eqref{eq:second_eq}
\begin{eqnarray}
\Pr\left\{\hspace{-0.2cm}\sup_{(\bm{u},\bm{v}) \in \mathcal{C}_{c,r}} \hspace{-0.3cm}\bm{u}^T \bm{P}_k \bm{v} - \bm{u}^* \bm{P}_k \bm{v}^* \geq \epsilon  \right\}  
\leq  \Pr\left\{\bigcup_{i=1}^{\mathcal{N}_{\epsilon}} \mathcal{B}_i\geq g_\epsilon \right\} \leq \sum_{i=1}^{\mathcal{N}_{\epsilon}} \Pr\left\{\mathcal{B}_i \geq \frac{\epsilon}{4}\right\}, \nonumber
\end{eqnarray}

where $g_\epsilon := \frac{\epsilon}{2} - \epsilon \norm{\widehat{\Delta P}^{(t)}}_{op}/4 \leq \frac{\epsilon}{4}$, using $\norm{\widehat{\Delta P}^{(t)}}_{op} \leq 1$, and $\mathcal{B}_i := \abs{\bm{x}_i \widehat{\Delta P}^{(t)} \bm{y}_i}$. Using the fact that $\abs{\bm{x}_i \widehat{\Delta P}^{(t)} \bm{y}_i} \leq \kappa \max_{j} \norm{\bm{x}_j}_1 \max_{i} \norm{\bm{y}_i}_1  \norm{\widehat{\Delta P}^{(t)}}_F \leq \kappa rc \norm{\widehat{\Delta P}^{(t)}}_F$, the above can be further bounded to get $\mathcal{N}_{\epsilon} \Pr\left\{\norm{\widehat{\Delta P}^{(t)}}_F \geq \frac{\epsilon}{4 \kappa r c}\right\}$. This completes the proof.

\section{Proof of Theorem \ref{thm:point_est_guarant_1bs}} \label{appendix:b}
Consider the following 
\begin{eqnarray} \label{eq:bound_2}
\Pr\left\{\norm{\widehat{\Delta P}^{(t)}}^2_F \geq \gamma\right\} \leq \Pr\left\{\max_{k,l}(p_{kl} - \hat{p}^{(t)}_{kl})^2 \hspace{-0.2cm}\geq \frac{\gamma}{F^2} \right\} \leq  F^2 \Pr\left\{ (p_{kl} - \hat{p}^{(t)}_{kl})^2 \geq \frac{\gamma}{F^2} \right\}, 
\end{eqnarray}
where $\gamma := \frac{\epsilon}{4 \kappa r c}$, and the second inequality above follows from the union bound. Conditioning on $V_{ls} := \sum_{s=1}^{t} v_l^{(s-1)} = m$, there are $Nm$ i.i.d. samples available to estimate $p_{kl}$. Using Hoeffdings inequality
\begin{eqnarray} \nonumber
\mathbb{E}\Pr\left\{ (p_{kl} - \hat{p}^{(t)}_{kl})^2 \geq \frac{\gamma}{F^2} \left|\right. V_{ls} = m\right\} 
\hspace{-0.3cm}&\leq& \hspace{-0.3cm} 2 \mathbb{E} \exp\left\{-\frac{2Nm \gamma}{F^2}  \right\}.
\end{eqnarray}
Since $V_{ls}$ is a binomial random variable with parameter $q$, the above average with respect to $V_{ls}$ becomes
\begin{equation}
2 \left(1- q \left( 1 - \exp\{-2N \gamma/F^2\}\right) \right)^t. 
\end{equation}
The following bound on the left hand side of \eqref{eq:thm_point_est_bound} can be obtained using the above in \eqref{eq:bound_2}, and substituting it in \eqref{eq:thm_point_est_bound}
\begin{equation}
2 \mathcal{N}_{\epsilon} F^2 \left(1- q \left( 1 - \exp\{-2N \gamma/F^2\}\right) \right)^t. 
\end{equation}
An upper bound on the above can be obtained by using $1 - x \leq e^{-x}$. Using the resulting bound, $\Pr\left\{\sup_{(\bm{u},\bm{v}) \in \mathcal{C}_{c,r}} \bm{u}^T \bm{P}_k \bm{v} - \bm{u}^* \bm{P}_k \bm{v}^* \geq \epsilon\right\} < \delta$ provided $t$ satisfies the bound in the theorem.
\section{Genie Aided Regret Analysis: Heuristics for Two sBSs Case} \label{sec:theoretical_guarant_point}
Consider the instantaneous regret given by $\texttt{Reg}_k(t) := (\bm{u}_{k,t}^*)^T \mathbf{P}_k \bm{v}_{k,t}^* \geq  \sup_{(\bm{u},\bm{v}) \in \mathcal{C}_{c,r}} \bm{u}_{k,t}^T \mathbf{P}_k \bm{v}_{k,t}$ at time $t$. Using the union bound, we can write
\begin{eqnarray}
\Pr\left\{\frac{1}{T} \sum_{t = 1}^{T}\texttt{Reg}_k(t) \geq \frac{1}{T}\sum_{t=1}^{T}\epsilon_t\right\} \leq \sum_{t=1}^T \Pr\left\{ \texttt{Reg}_k(t) \geq \epsilon_t\right\}, \label{eq:point_est_bound}
\end{eqnarray}
where $\epsilon_t > 0$. From Theorem \ref{thm:point_est_guarant_1bs}, it follows that for any $\epsilon_t > 0$, we have $\Pr\left\{ \texttt{Reg}_k(t) \leq \epsilon_t\right\} \geq 1- \delta$ provided \eqref{eq:bound_ontime_single_BS}. By choosing $\delta = \frac{1}{T^2}$, the approximation $e^{-x} \approx 1-x$ for small $x$, and $|\mathcal{N_\epsilon}| \lessapprox 1/\epsilon^F$ in Theorem \ref{thm:point_est_guarant_1bs}, we get $\Pr\left\{ \texttt{Reg}_k(t) \leq \epsilon_t\right\} \geq 1- \frac{1}{T^2}$ provided
\begin{equation}
t \gtrapprox \frac{8 \kappa^2 F^2 c^2 r^2}{q N\epsilon_t^3}\log \frac{2F^2T^2}{\epsilon_t^F}, \label{eq:time_point_bound}
\end{equation}
where $\gtrapprox$ is used to denote ``approximately greater than or equal to". Assuming $\epsilon_t <1 $ and using $\log x \approx x$ for small $x$, we have $\log \frac{2F^2T^2}{\epsilon_t^F} = \log 2F^2T^2 + F \log\frac{1}{\epsilon_t} \leq \frac{(\log2F^2T^2 + F)}{\epsilon_t}$.\footnote{The case of $\epsilon_t >1$ can be handled in a similar fashion, and hence ignored.} Now, we can use \eqref{eq:time_point_bound} to write $\epsilon_t$ in terms of $t$ to get
\begin{equation}
\epsilon_t \gtrapprox \sqrt[3]{\frac{8\kappa^2F^2c^2r^2 (\log2F^2T^2 + F)}{qNt}}. 
\end{equation}
Note that by choosing large enough $t$, the above can be made less than one. Since we are looking for order result, this does not change the final result. In other words, with a probability of at least $1- \frac{1}{T^2}$, $\texttt{Reg}_k(t) \leq \sqrt[3]{\frac{8\kappa^2F^2c^2r^2 (\log2F^2T^2 + F)}{qNt}}$. Using this result in \eqref{eq:point_est_bound}, we get the following result. With a probability of at least $1-\frac{1}{T}$, 
\begin{eqnarray}
\texttt{Reg}_{k,T} &\lessapprox& \sqrt[3]{\frac{8\kappa^2F^2c^2r^2 (\log2F^2T^2 + F)}{qN}}\sum_{t=1}^T \frac{1}{t^{1/3}}  \nonumber \\
&=& \mathcal{O}(T^{2/3} \sqrt{\log T}).
\end{eqnarray}
Thus, the above shows that the regret achieved grows sub-linearly with time, and hence (genie aided) achieves a zero asymptotic average regret. 
\section{Regret Analysis for Two sBS: Heuristics} \label{app:2sbs_heuristics}
The analysis here is very similar to the analysis of single BS case. We repeat some of the analysis for the sake of clarity and completeness. Let the instantaneous regret at the BS $k$ at time $t$ is given by $\texttt{Reg}_k(t) := (\bm{u}_{k,t}^*)^T \mathbf{P}_k \bm{v}_{k,t}^* \geq  \sup_{(\bm{u},\bm{v}) \in \mathcal{C}_{c,r}} \bm{u}_{k,t}^T \mathbf{P}_k \bm{v}_{k,t}$ at time $t$. The union bound results in 
\begin{eqnarray}
\Pr\left\{\frac{1}{T} \sum_{t = 1}^{T}\texttt{Reg}_k(t) \geq \frac{1}{T}\sum_{t=1}^{T}\epsilon_{k,t}\right\} \leq \sum_{t=1}^T \Pr\left\{ \texttt{Reg}_k(t) \geq \epsilon_{k,T}\right\}, \label{eq:point_est_bound_2bs}
\end{eqnarray}
where $\epsilon_{k,t} > 0$ and $\epsilon_{k,T} = \frac{1}{T}\sum_{t=1}^T \epsilon_{k,t}$. From Theorem \ref{thm:point_est_guarant_1bs}, it follows that for any $\epsilon_{k,t} > 0$, we have $\Pr\left\{ \texttt{Reg}_k(t) \leq \epsilon_{k,t}\right\} \geq 1- \delta$ provided \eqref{eq:t_bound} is satisfied. By choosing $\delta = \frac{1}{T^2}$, assuming $\epsilon_{k,t} <1 $, using the approximations $e^{-x} \approx 1-x$ for small $x$, and $|\mathcal{N_\epsilon}| \lessapprox 1/\epsilon^F$, we have $\log \frac{2F^2T^2}{\epsilon_{k,t}^F} \leq \frac{(\log2F^2T^2 + F)}{\epsilon_{k,t}}$. Using this in Theorem \ref{thm:point_est_guarant_1bs}, we get $\Pr\left\{ \texttt{Reg}_k(t) \leq \epsilon_{k,t}\right\} \geq 1- \frac{1}{T^2}$ provided
\begin{equation}
\tau\bigg(\frac{\epsilon_{k,t}}{\lambda_k}, \frac{\delta}{2}\bigg) \gtrapprox \frac{8 \kappa^2 F^2 c^2 r^2 \lambda_k^3(\log 4F^2T^2 + F)}{q N\epsilon_{k,t}^3}, \label{eq:time_point_bound_2sbs}
\end{equation}
where $\gtrapprox$ is used to denote ``approximately greater than or equal to".  
Using the above in \eqref{eq:t_bound}, we get 
\begin{eqnarray}
t \gtrapprox \max\Bigg\{\frac{8 \kappa^2 F^2 c^2 r^2 \lambda_k^3 (\log 4F^2T^2 + F)}{q N\epsilon_{k,t}^3}, \frac{8 \kappa^2 F^2 c^2 r^2 (1-\lambda_k)^3(\log 4F^2T^2 + F)}{q N\epsilon_{k,t}^3}\Bigg\} .\nonumber
\end{eqnarray}
By rearranging and summing over $t$, the error can be written as follows
\begin{eqnarray}
\epsilon_{k,t} \lessapprox \frac{1}{\sqrt[3]{t}}\max \Bigg\{ \Theta \lambda_k, \Theta (1- \lambda_k)  \Bigg\},  \nonumber
\end{eqnarray}
where $\Theta = \sqrt[3]{\frac{8 \kappa^2 F^2 c^2 r^2(\log 4F^2T^2 + F)}{q N}} $. Using this in the place of $\epsilon_{k,t}$ in the above theorem, and summing over $t$, we get with a probability of at least $1-\frac{1}{T}$, the following holds for BS $k$
\begin{eqnarray}
\texttt{Reg}_{k,T} \lessapprox \max \Bigg\{ \Theta \lambda_k  , \Theta(1-\lambda_k)  \Bigg\} \frac{1}{\sqrt[3]{T}}\sum_{t =1}^T 1 + 2T(1 - \lambda_k)\mathcal{V}_{12}, \nonumber
\end{eqnarray}
where $\mathcal{V}_{12}: = \sup_{(\bm{u},\bm{v}) \in \mathcal{C}_{c,r}} \abs{\bm{u}^T (\textbf{P}_2 - \textbf{P}_1) \bm{v}}$ and $\Theta = \sqrt[3]{\frac{8 \kappa^2 F^2 c^2 r^2(\log 4F^2T^2 + F)}{q N}} $. This completes the approximate analysis.  
\section{Regret Analysis for Multiple sBS: Heuristics} \label{app:regret_multisbs_heuristics_genieaided}
The analysis here is again very similar to the analysis of single sBS case.
From Theorem \ref{thm:point_est_guarant_1bs}, it follows that for any $\epsilon_{k,t} > 0$, we have $\Pr\left\{ \texttt{Reg}_k(t) \leq \epsilon_{k,t}\right\} \geq 1- \delta$, where $\epsilon_{k,t} > 0$ and $\epsilon_k = \frac{1}{T}\sum_{t=1}^T \epsilon_{k,t}$. By choosing $\delta = \frac{1}{T^2}$, assuming $\epsilon_{k,t} <1 $, using the approximation $e^{-x} \approx 1-x$ for small $x$,  and $|\mathcal{N_\epsilon}| \lessapprox 1/\epsilon^F$, we have $\log \frac{2F^2T^2}{\epsilon_{k,t}^F} \leq \frac{(\log2F^2T^2 + F)}{\epsilon_{k,t}}$. Using this in Theorem \ref{thm:point_est_guarant_1bs}, we get $\Pr\left\{ \texttt{Reg}_k(t) \leq \epsilon_{k,t}\right\} \geq 1- \frac{1}{T^2}$ provided
\begin{equation}
\tau\bigg(\frac{\epsilon_1}{\lambda_1^{(k)}}, \frac{\delta}{M}\bigg) \gtrapprox \frac{8 \kappa^2 F^2 c^2 r^2 \lambda_k^3(\log 4F^2T^2 + F)}{q N\epsilon_{k,t}^3}, \label{eq:time_point_bound_msbs}
\end{equation}
where $\gtrapprox$ is used to denote ``approximately greater than or equal to".  
Using the above in \eqref{eq:t_bound_msbs}, we get 
\begin{eqnarray}
t \gtrapprox \frac{\Theta^3}{\epsilon_{k,t}^3} \max \left\{(\lambda_1^{(k)})^3, (\lambda_2^{(k)})^3, \ldots,  (\lambda_M^{(k)})^3 \right\}, \nonumber
\end{eqnarray}
where, $\Theta = \sqrt[3]{\frac{8 \kappa^2 F^2 c^2 r^2(\log F^2T^2 + F)}{q N}  }$.
From the above, it is clear that the waiting time $t$ scales as the square of $F, c$ and $r$, and is inversely proportional to the error $\epsilon_{k,t}^3$. By rearranging and summing over $t$, the error can be written as $\epsilon_{k,t} \lessapprox \frac{1}{\sqrt[3]{t}}\max \Bigg\{ \Theta \lambda_1^{(k)}, \ldots, \Theta \lambda_M^{(k)}  \Bigg\} $. Using this in the place of $\epsilon_{k,t}$ in the above theorem, and summing over $t$, we get with a probability of at least $1-\frac{1}{T}$ the following result on the regret for BS $k$ holds
\begin{eqnarray}
\texttt{Reg}_{k,T} \lessapprox \frac{\Theta M^2}{\sqrt[3]{T}}\max \Bigg\{  \lambda_1^{(k)}  , \ldots, \lambda_M^{(k)}  \Bigg\} T+ M^2T\big\{(1 - \lambda_1^{(k)}) \mathcal{D}_{1} - \ldots,  -\lambda_M^{(k)} \mathcal{D}_M\big\}. \nonumber 
\end{eqnarray}	
\section{Proof of Theorem \ref{thm:bay_guarant_1bs}} \label{app:regret_oneBS}
Similar to the proof of Theorem \ref{thm:bound_first_general}, from \cite{bousquet2003introduction}, it follows that at time $t$, the performance gap of the proposed algorithm with respect to the optimal is given by 
\begin{eqnarray}
\bm{u}^* \bm{P}_k \bm{v}^* - \bm{u}_t^T \bm{P}_k \bm{v}_t &\leq& 2 \sup_{(\bm{u},\bm{v}) \in \mathcal{C}_{c,r}} \abs{\bm{u}^T \widehat{\Delta P}^{(t)} \bm{v}}. \nonumber 
\end{eqnarray}
Summing the above over all $t$, we get
\begin{eqnarray}
T\bm{u}^* \bm{P}_k \bm{v}^* - \sum_t\bm{u}^T_t \bm{P}_k \bm{v}_t\hspace{-0.3cm} &\leq& \hspace{-0.3cm}2 \sum_t \sup_{(\bm{u},\bm{v}) \in \mathcal{C}_{c,r}} \abs{\bm{u}^T \widehat{\Delta P}^{(t)} \bm{v}}. \nonumber
\end{eqnarray}

For a given $\epsilon$, the above implies that 
\begin{eqnarray} \label{eq:regret_bound}
\Pr \bigg\{ T\bm{u}^* \bm{P}_k \bm{v}^* -  \sum_t\bm{u}^T_t \bm{P}_k \bm{v}_t \geq \frac{\epsilon}{2}  \bigg\} 
&\leq& \hspace{-0.3cm}\Pr \bigg\{ \sum_t \sup_{(\bm{u},\bm{v}) \in \mathcal{C}_{c,r}} \abs{\bm{u}^T \widehat{\Delta P}^{(t)} \bm{v}} \geq \frac{\epsilon}{2} \bigg\} \nonumber \\
&= &\hspace{-0.3cm} \Pr\bigg\{ \sum_t Y_t \geq \frac{\epsilon}{2} - \sum_t \mathbb{E}\left[ \sup_{(\bm{u},\bm{v}) \in \mathcal{C}_{c,r}}\hspace{-0.3cm} \abs{\bm{u}^T \widehat{\Delta P}^{(t)} \bm{v}}\right] \hspace{-0.2cm}\bigg\}, 
\end{eqnarray}
where $Y_t =  \sup_{(u,v) \in \mathcal{N}_{\epsilon}} \abs{\bm{u} \widehat{\Delta P}^{(t)} \bm{v}} - 
\mathbb{E}\bigg[ \sup_{(u,v) \in \mathcal{N}_{\epsilon}}\abs{\bm{u} \widehat{\Delta P}^{(t)} \bm{v}} \bigg]$
is a martingale difference, i.e.,  $\mathbb{E}\{Y_t\} = 0$, and $\mathcal{N}_\epsilon$ is the covering set of $\mathcal{C}_{c,r}$ as in Definition I. By Azuma's inequality, we have
\begin{equation} \label{eq:bound}
\Pr \bigg\{ \sum_t Y_t > \frac{\epsilon}{2} \bigg\} \leq \exp\bigg\{ \frac{-\epsilon^2}{2 (4rc)^2}\bigg\}.
\end{equation}
The above follows due to the fact that $\abs{Y_t} \leq 4rc$, which is explained below:
\begin{eqnarray}
Y_t &\leq& \hspace{-0.3cm} \sup_{(\bm{u},\bm{v}) \in \mathcal{N}_{\epsilon}} \abs{\bm{u}^T \widehat{\Delta P}^{(t)} \bm{v}} + \mathbb{E} \bigg[ \sup_{(\bm{u},\bm{v}) \in \mathcal{N}_{\epsilon}} \abs{\bm{u}^T \widehat{\Delta P}^{(t)} \bm{v}}  \bigg]\nonumber \\
&\leq& \sup_{(\bm{u},\bm{v}) \in \mathcal{N}_{\epsilon}} \abs{\bm{u}^T {P} \bm{v}} + \sup_{(\bm{u},\bm{v}) \in \mathcal{N}_{\epsilon}} \abs{\bm{u}^T \widehat{P}^{(t)} \bm{v}} + \mathbb{E} \bigg[ \sup_{(\bm{u},\bm{v}) \in \mathcal{N}_{\epsilon}} \abs{\bm{u}^T P \bm{v}}  \bigg] + \mathbb{E} \bigg[ \sup_{(\bm{u},\bm{v}) \in \mathcal{N}_{\epsilon}} \abs{\bm{u}^T \widehat{P}^{(t)} \bm{v}}  \bigg] \nonumber \\
&\leq&  4rc,
\end{eqnarray}
which follows from $\abs{Y_t} \leq 4rc$ and $\sup_{(\bm{u},\bm{v}) \in \mathcal{C}_{c,r}} \abs{\bm{u}^T {P} \bm{v}} \leq rc$.
Thus it follows from \eqref{eq:bound}, $\Pr \bigg\{ \sum_t Y_t > \frac{\epsilon}{2} \bigg\} \leq \delta$ if $\epsilon \geq 32r^2c^2T \log(1/\delta)$. Using this definition of $\abs{Y_t}$, it follows that with a probability of at most $\delta$, we have
\begin{eqnarray}
\sum_t\sup_{(u,v) \in \mathcal{N}_{\epsilon}} \abs{\bm{u} \widehat{\Delta P}^{(t)} \bm{v}} \geq 2 \sum_t \mathbb{E} \bigg[ \sup_{(\bm{u},\bm{v}) \in \mathcal{C}_{c,r}} \abs{\bm{u}^T \widehat{\Delta P}^{(t)} \bm{v}}  \bigg] + \sqrt{128r^2c^2T\log(1/\delta)}. \nonumber
\end{eqnarray}
Choosing $\epsilon = 32r^2c^2T \log(1/\delta)$ in \eqref{eq:regret_bound}, the following bound for regret is satisfied with a probability of at least $1 - \delta$
\begin{eqnarray} \label{eq:reg_bound}
\bm{u}^* \bm{P}_k \bm{v}^* - \frac{1}{T}\sum_{t = 1}^{T} \bm{u}^T_t \bm{P}_k \bm{v}_t < \frac{2}{T} \sum_t \mathbb{E} \bigg[ \hat{\Delta}_t \bigg]  +   \sqrt{\frac{128r^2c^2 \log(1/\delta)}{T}}, \nonumber
\end{eqnarray}
where $\hat{\Delta}_t := \sup_{(\bm u,\bm v) \in \mathcal{N}_{\epsilon}} \abs{\bm{u}^T \widehat{\Delta P}^{(t)} \bm{v}}$. Now, it remains to bound the first term on the right hand side above. for a given $\psi_t > 0$ (to be chosen later), using the total expectation rule, we get
\begin{eqnarray}
\sum_t \mathbb{E}  \hat{\Delta}_t &\leq&  \sum_t \mathbb{E} \bigg[ \hat{\Delta}_t \left|\right. \hat{\Delta}_t  > \psi_t  \bigg] \Pr \bigg\{\hat{\Delta}_t  > \psi_t  \bigg\} + \sum_t \psi_t \nonumber \\
&\leq&  rc \max_{k,l} p_{kl} \sum_t \Pr \bigg\{ \hat{\Delta}_t  > \psi_t \bigg\} + \sum_t\psi_t, \nonumber 
\end{eqnarray}
where the first inequality above follows by using the bound $\sup_{(\bm{u},\bm{v}) \in \mathcal{C}_{c,r}} \abs{\bm{u}^T \widehat{\Delta P}^{(t)} \bm{v}} \leq \psi_t$. Since $\sup_{(\bm{u},\bm{v}) \in \mathcal{C}_{c,r}} \abs{\bm{u}^T \widehat{\Delta P}^{(t)} \bm{v}} > \sup_{(\bm{u},\bm{v}) \in \mathcal{N}_{\epsilon}} \abs{\bm{u}^T \widehat{\Delta P}^{(t)} \bm{v}}$, we get
\begin{eqnarray} \label{eq:exp_bound}
\sum_t\mathbb{E} \bigg[ \sup_{(u,v) \in \mathcal{N}_{\epsilon}} \abs{\bm{u}^T \widehat{\Delta P}^{(t)} \bm{v}} \bigg]  \leq rc |\mathcal{N}_\epsilon| \max_{k,l} p_{kl}  
\sum_t \Pr \bigg\{ \sup_{(u,v) \in \mathcal{N}_{\epsilon}} \abs{\bm{u}^T \widehat{\Delta P}^{(t)} \bm{v}} > \psi_t \bigg\} + \sum_t\psi_t.
\end{eqnarray}
Now, consider
\begin{eqnarray} \label{eq:prob_bound}
\sum_t\Pr \bigg\{ \sup_{(u,v) \in \mathcal{N}_{\epsilon}} \abs{\bm{u}^T \widehat{\Delta P}^{(t)} \bm{v}} > \psi_t \bigg\} 
&\stackrel{(a)}{\leq}& \sum_t \Pr \bigg\{ \sum_{i,j=1}^{\abs{\mathcal{N_\epsilon}}} \bm{u}_i^T \widehat{\Delta P}^{(t)} \bm{v}_j > \psi_{t}\bigg\} \nonumber \\
&\leq & \sum_t \Pr \bigg\{ \sum_{i, j=1}^{\abs{\mathcal{N_\epsilon}}}\sum_{l=1}^{\abs{\mathcal{N_\epsilon}}} u_{ij}\widehat{\Delta P}^{(t)T}_{j} v_l> \psi_t\bigg\} \nonumber \\
&\leq & \sum_t \Pr \bigg\{ e^{s\sum_{i,j=1}^{\abs{\mathcal{N_\epsilon}}} u_{ij}^T X_j} > e^{s\psi_t}\bigg\} \nonumber \\
& \stackrel{(b)}{\leq}& \sum_t  e^{-s\psi_t} \mathbb{E}[e^{s\sum_{i,j=1}^{\abs{\mathcal{N_\epsilon}}} u_{ij}^T X_j}]
\end{eqnarray}
where $X_j := \sum_{l=1}^{\abs{\mathcal{N_\epsilon}}}v_l \widehat{\Delta P}^{(t)}_{lj}$. In the above, $(a)$ follows from the covering argument, and $(b)$ follows from the Chernoff bound. From \cite{Marchal_2017}, using the optimal proxy variance, we get the following bound  
\begin{eqnarray}\label{eq:exp_bound_2}
\sum_t \Pr \bigg\{ \sup_{(u,v) \in \mathcal{N}_{\epsilon}} \abs{\bm{u}^T \widehat{\Delta P}^{(t)} \bm{v}} > \psi_t \bigg\} \leq\sum_t    \exp\left\{-s\psi_t+ s^2\sum_{i,j,l=1}^{\abs{\mathcal{N_\epsilon}}} \frac{\norm{u_{il}}^2 \norm{v_j}^2\sigma_j^2}{2}\right\}, 
\end{eqnarray}
where an upper bound on $\sigma_l$ (see \cite{Marchal_2017}) is given by $\sigma_j \leq \frac{1}{4(\sum_i \alpha_{ij}^{(t)} + 1)}$ and $\alpha_{ij}^{(t)} = \sum_{q=1}^{t-1}d_i^{(q)}v_j^{(q-1)}$. Optimizing the exponent in \eqref{eq:exp_bound_2}, 
the optimal $s^* = \frac{\psi_t}{\sum_{i,j,l=1}^{\abs{\mathcal{N_\epsilon}}} \norm{u_{il}}^2 \norm{v}_F^2\sigma_j^2}$.
Further, $\norm{u_{il}}^2 \leq \norm{u_{il}} \leq c$ and $\norm{v}_F^2 \leq r |\mathcal{N}_{\epsilon}|^2$. Using these bounds, and the bound on $\sigma_l$ above, \eqref{eq:exp_bound_2} can be written as follows
\begin{eqnarray}
\sum_t\Pr \bigg\{ \sup_{(u,v) \in \mathcal{N}_{\epsilon}} \abs{\bm{u}^T \widehat{\Delta P}^{(t)} \bm{v}} > \psi_t \bigg\} \leq \sum_t\exp\bigg\{\frac{-8\psi_t^2}{cr\abs{\mathcal{N_\epsilon}}^2 \bar{\sigma}^2(t)}  \bigg\},
\end{eqnarray}
where $\bar{\sigma}^2(t):= \left[\sum_{j} \frac{1}{\left(\sum_i \alpha_{ij}^{(t)} +1\right)^2}\right]$. Substituting the above in \eqref{eq:exp_bound} results in
\begin{eqnarray}
\hspace{-0.3cm}\sum_t\mathbb{E} \bigg[ \sup_{(u,v) \in \mathcal{N}_{\epsilon}} \abs{\bm{u}^T \widehat{\Delta P}^{(t)} \bm{v}} \bigg]  \leq rc \max_{ij} p_{ij} |N_\epsilon|\sum_t\exp\bigg\{\frac{-8\psi_t^2}{cr\abs{\mathcal{N_\epsilon}}^2 \bar{\sigma}^2(t)}  \bigg\} + \sum_t\psi_t.
\end{eqnarray}
Thus, using the above, the regret can be written as
\begin{eqnarray}
\texttt{Reg}_T \leq {2rc\max_{ij} p_{ij} |\mathcal{N}_\epsilon|}\sum_t\exp\bigg\{\frac{-8\psi_t^2}{cr\abs{\mathcal{N}_\epsilon}^2 \bar{\sigma}^2(t)}  \bigg\} + {2\sum_t\psi_t} + \sqrt{128r^2c^2T\log(1/\delta)}.
\end{eqnarray}
Now, the proof is complete by choosing $\delta = 1/T$.
\section{Proof of Theorem \ref{thm:point_est_guarant_2bs}} \label{appendix:d}
The analysis is done only for the first sBS as the analysis for the second sBS is similar. As in \eqref{eq:first_eq}, since $\Pr\bigg\{(\bm{u}_{k,t}^*)^T \textbf{P}_k \bm{v}_{k,t}^* \geq \sup_{(\bm{u},\bm{v}) \in \mathcal{C}_{c,r}} \bm{u}^T \hat{\bm{Q}}^{(t)}_k \bm{v} - \epsilon \bigg\} \leq \Pr\left\{2\sup_{(\bm{u},\bm{v}) \in \mathcal{C}_{c,r}} \abs{\bm{u}^T \widehat{\Delta P}^{(t)} \bm{v}} > \epsilon\right\}$, it is sufficient to consider the following 
\begin{eqnarray} \label{eq:first_eq_twobs}
&&\sup_{(\bm{u},\bm{v}) \in \mathcal{C}_{c,r}} \abs{\bm{u}^T \widehat{\Delta P}^{(t)} \bm{v}}  =  \sup_{(\bm{u},\bm{v}) \in \mathcal{C}_{c,r}} \Bigg|\bm{u}^T (\textbf{P}_1 - \hat{\bm{Q}}^{(t)}_1 ) \bm{v} \Bigg|\nonumber \\
&=&  \sup_{(\bm{u},\bm{v}) \in \mathcal{C}_{c,r}} \Bigg|\lambda_1 \bm{u}^T \widehat{\Delta P}_1^{(t)} \bm{v} + (1 - \lambda_1)\bm{u}^T \widehat{\Delta P}_2^{(t)} \bm{v} + (1 - \lambda_1)\bm{u}^T (\textbf{P}_1 -  \textbf{P}_2) \bm{v}\Bigg|\nonumber \\
&\leq& \lambda_1 \mathcal{V}_1  + (1-\lambda_1) \mathcal{V}_2 + (1 - \lambda_1)\mathcal{V}_{12},
\end{eqnarray} 
where $\mathcal{V}_1  := \sup_{(\bm{u},\bm{v}) \in \mathcal{C}_{c,r}} \abs{\bm{u}^T \widehat{\Delta P_1}^{(t)} \bm{v}}$, $\mathcal{V}_2 : = \sup_{(\bm{u},\bm{v}) \in \mathcal{C}_{c,r}} \abs{\bm{u}^T \widehat{\Delta P_2}^{(t)} \bm{v}}$, and \nonumber \\ $\mathcal{V}_{12}: = \sup_{(\bm{u},\bm{v}) \in \mathcal{C}_{c,r}} \abs{\bm{u}^T (\textbf{P}_2 - \textbf{P}_1) \bm{v}}$. Here, $\widehat{\Delta P}^{(t)} := \textbf{P}_1 - \hat{Q}_1^{(t)}$, $\widehat{\Delta P_k}^{(t)} := \textbf{P}_k - \hat{\textbf{P}}_k^{(t)}$, $k=1,2$. Using the union bound, we get the following
\begin{eqnarray}
\Pr\{\lambda_1 \mathcal{V}_1  + (1-\lambda_1) \mathcal{V}_2  > \epsilon_{1} \}  \leq \Pr\left\{\mathcal{V}_1   > \frac{\epsilon_{1}}{\lambda_1 }\right\} + \Pr\left\{ \mathcal{V}_2  > \frac{\epsilon_{1}}{(1-\lambda_1)}\right\},\nonumber
\end{eqnarray}
where $\epsilon_{1} := \epsilon/2 - (1 - \lambda_1) \mathcal{V}_{12}$.
Using results from Theorem \ref{thm:point_est_guarant_1bs} to each of the above term with $\epsilon$ replaced by $\epsilon_{1}/\lambda_1$ and $\frac{\epsilon_{1}}{(1-\lambda_1)}$ with $\delta$ replaced by $\delta/2$ proves the theorem. 
\section{Proof of Theorem \ref{thm:bay_est_guarant_2bs}} \label{appendix:e}
Similar to the proof provided of Theorem 4.1, the analysis is done only for the first sBS using Bayesian estimate. 
\begin{eqnarray} \label{eq:first_eq_twobs}
T\bm{u}^* \textbf{P}_1 \bm{v}^* - \sum_t\bm{u}^T_t \hat{\bm{Q}}^{(t)}_k \bm{v}_t \leq \lambda_1 \mathcal{U}_1  + (1-\lambda_1) \mathcal{U}_2 + (1 - \lambda_1)\mathcal{U}_{12}, \nonumber
\end{eqnarray} 
where $\mathcal{U}_1  := \sum_t \sup_{(\bm{u},\bm{v}) \in \mathcal{C}_{c,r}} \abs{\bm{u}^T \widehat{\Delta P_1}^{(t)} \bm{v}}$, $\mathcal{U}_2 : = \sum_t \sup_{(\bm{u},\bm{v}) \in \mathcal{C}_{c,r}} \abs{\bm{u}^T \widehat{\Delta P_2}^{(t)} \bm{v}}$, and \nonumber \\ $\mathcal{U}_{12}: = \sum_t \sup_{(\bm{u},\bm{v}) \in \mathcal{C}_{c,r}} \abs{\bm{u}^T (\textbf{P}_1 - \textbf{P}_2) \bm{v}}$. Here, $\widehat{\Delta P_k}^{(t)} := \textbf{P}_k - \hat{\textbf{P}}_k^{(t)}$, $k=1,2$. Consider the following 
\begin{eqnarray}
\Pr \{T\bm{u}^* \textbf{P}_1 \bm{v}^* - \sum_t\bm{u}^T_t \hat{\bm{Q}}^{(t)}_1 \bm{v}_t \geq \epsilon\}&\leq& \Pr \{\lambda_1 \mathcal{U}_1  + (1-\lambda_1) \mathcal{U}_2 + (1 - \lambda_1)\mathcal{U}_{12} \geq \epsilon\} \nonumber \\
&\leq&  \Pr\left\{\mathcal{U}_1   > \frac{\epsilon_1}{\lambda_1 }\right\} + \Pr\left\{ \mathcal{U}_2  > \frac{\epsilon_1}{(1-\lambda_1)}\right\}, \nonumber
\end{eqnarray}
where $\epsilon_1 := \epsilon/2 - (1 - \lambda_1) \mathcal{U}_{12}$. Using results from Theorem \ref{thm:bay_guarant_1bs} to each of the above term with $\epsilon$ replaced by $2\epsilon_1/\lambda_1$ and $\frac{2\epsilon_1}{(1-\lambda_1)}$ and $\delta$ replaced by $\delta/2$, we get the desired result. 
\section{Proof of Theorem \ref{thm:point_est_guarant_mbs}} \label{appendix:f}
The proof for multiple sBSs is a generalization of two sBSs and the analysis is done only for the first sBS, the rest of the sBSs are similar.
\begin{eqnarray} \label{eq:first_eq_multiplebs}
&&\sup_{(\bm{u},\bm{v}) \in \mathcal{C}_{c,r}} \abs{\bm{u}^T \widehat{\Delta P}^{(t)} \bm{v}} = \sup_{(\bm{u},\bm{v}) \in \mathcal{C}_{c,r}} \abs{\bm{u}^T (\textbf{P}_1 - \lambda^{(k)}_1 \hat{\textbf{P}}_1^{(t)}, \ldots - \lambda^{(k)}_M \hat{\textbf{P}}_M^{(t)}) \bm{v}} \nonumber \\
&=&   \sup_{(\bm{u},\bm{v}) \in \mathcal{C}_{c,r}} \bigg|\sum_{i=1}^M \lambda^{(k)}_i\bm{u}^T (\textbf{P}_i - \hat{\textbf{P}}_i^{(t)}) \bm{v} + (1 - \lambda_1^{(k)})\bigg\{\sup_{(\bm{u},\bm{v}) \in \mathcal{C}_{c,r}} \abs{\bm{u}^T \textbf{P}_1 \bm{v}}\bigg\}-   \sum_{i=1}^M\lambda_i^{(k)} \bigg\{\sup_{(\bm{u},\bm{v}) \in \mathcal{C}_{c,r}} \abs{\bm{u}^T P_i \bm{v}}\bigg\}\bigg|\nonumber \\
&\leq& \sum_{i=1}^M \lambda_i^{(k)} \mathcal{V}_i +  (1 - \lambda_1^{(k)})\mathcal{D}_1 - \sum_{i=2}^M \lambda_i^{(k)} \mathcal{D}_i,\nonumber
\end{eqnarray} 
where $\mathcal{V}_l := \sup_{(\bm{u},\bm{v}) \in \mathcal{C}_{c,r}} \abs{\bm{u}^T \widehat{\Delta P_l}^{(t)} \bm{v}}$,  $\mathcal{D}_{l}: = \sup_{(\bm{u},\bm{v}) \in \mathcal{C}_{c,r}} \abs{\bm{u}^T \textbf{P}_l \bm{v}}$, $l=1,2,\ldots,M$. Here, $\widehat{\Delta P_k}^{(t)} := \textbf{P}_1 - \hat{\textbf{Q}}_1^{(t)}$, $\widehat{\Delta P_k}^{(t)} := \textbf{P}_k - \hat{\textbf{P}}_k^{(t)}$, $k=1,2, \ldots, M$. 
Thus we can write the following
\begin{eqnarray}
\Pr\left\{\sum_{i=1}^M\lambda_i^{(k)} \mathcal{V}_i  > \epsilon^{'} \right\}  
\leq  \sum_{l=1}^M \Pr\left\{\mathcal{V}_l   > \frac{\epsilon_l}{\lambda_l^{(k)} }\right\},\nonumber 
\end{eqnarray}
where $\epsilon^{'} := \epsilon/M - (1 - \lambda_1^{(k)}) \mathcal{D}_{1} + \lambda_2^{(k)} \mathcal{D}_2 +, \ldots,  + \lambda_M^{(k)} \mathcal{D}_M$, and $\epsilon_{k} := \epsilon/M^2 - (1 - \lambda_1^{(k)}) \mathcal{D}_{1} + \lambda_2^{(k)} \mathcal{D}_2 +, \ldots,  + \lambda_M^{(k)} \mathcal{D}_M, \forall$ $k$ $= {1,2,\ldots, M}$. Using results from Theorem \ref{thm:point_est_guarant_1bs} to each of the above term with $\epsilon$ replaced by $\epsilon^{'}$ and $\epsilon_k$ and $\delta$ replaced by $\delta/M$ proves the theorem.
\section{Proof of Theorem \ref{thm:bays_est_guarant_mbs}} \label{appendix:g}
The proof for multiple sBSs is a generalization of two sBSs and the analysis is done only for the first sBS for Bayesian estimate, the rest of the sBSs are similar. First consider the following
\begin{eqnarray} \label{eq:first_eq_multiplebs}
T\bm{u}^* \textbf{P}_1 \bm{v}^* - \sum_t\bm{u}^T_t \textbf{P}_1 \bm{v}_t \leq \sum_{j=1}^M \lambda_j^{(k)} \mathcal{W}_j + (1 - \lambda_1^{(k)})\mathcal{I}_1 - \sum_{j=2}^M \lambda_j^{(k)} \mathcal{I}_j,
\end{eqnarray} 
where $\mathcal{W}_j := \sum_t \sup_{(\bm{u},\bm{v}) \in \mathcal{C}_{c,r}} \abs{\bm{u}^T \widehat{\Delta P_j}^{(t)} \bm{v}}$, $\mathcal{I}_{j}: = \sum_t \sup_{(\bm{u},\bm{v}) \in \mathcal{C}_{c,r}} \abs{\bm{u}^T \textbf{P}_j \bm{v}}$, $j=1,2,\ldots,M$. Here, $\widehat{\Delta P_k}^{(t)} := \textbf{P}_k - \hat{\textbf{P}}_k^{(t)}$, $k=1,2, \ldots, M$. 
Thus we can write the following:
\begin{eqnarray}
\Pr\{\lambda_1^{(k)} \mathcal{W}_1,.., + \lambda_M^{(k)} \mathcal{W}_M \geq \epsilon^{'} \} 
\leq  \sum_{l=1}^M \Pr\left\{\mathcal{W}_l   > \frac{\epsilon_l}{\lambda_l^{(k)} }\right\},\nonumber
\end{eqnarray}
where $\epsilon^{'} := \epsilon/M - (1 - \lambda_1^{(k)}) \mathcal{I}_{1} + \sum_{l=2}^M \lambda_l^{(k)} \mathcal{I}_k$, and $\epsilon_{k} := \epsilon/M^2 - (1 - \lambda_1^{(k)}) \mathcal{I}_{1} + \sum_{l=2}^M \lambda_l^{(k)} \mathcal{I}_k$, for all $k = \{1,2,\ldots, M\}$. Using results from Theorem \ref{thm:bay_guarant_1bs} to each of the above term with $\epsilon$ replaced by $\epsilon^{'}$ and $\epsilon_k$ and $\delta$ replaced by $\delta/M$, we get the regret bound described in the theorem.

\end{document}